\documentclass[aps,prb,twocolumn,amsmath,amssymb,superscriptaddress]{revtex4-1}
\usepackage{dcolumn}
\usepackage{bm}

\usepackage{color} 
\usepackage{ulem}
\usepackage{comment}
\usepackage{amsmath}
\usepackage{graphicx}
\usepackage{float}
\usepackage{subcaption}
\usepackage{tikz}
\usepackage{color}
\usepackage[colorlinks,bookmarks=false,citecolor=blue,linkcolor=red,urlcolor=blue]{hyperref}

\definecolor{darkred}{rgb}{0.7,0.0,0.0}

\definecolor{darkblue}{rgb}{0,0.02,0.45}

\definecolor{darkgreen}{rgb}{0.02,0.45,0.0}

\definecolor{violet}{rgb}{0.8,0.2,0.6}

\providecommand{\U}[1]{\protect\rule{.1in}{.1in}}

\begin{document}

\title{Anomalous Landau levels and quantum oscillation in rotation-invariant insulators}

\author{Jianlong Fu}
\affiliation{Department of Physics, Hong Kong University of Science and Technology, Clear Water Bay, Hong Kong, China}
\affiliation{Center for Theoretical Condensed Matter Physics, Hong Kong University of Science and Technology, Clear Water Bay, Hong Kong, China}

\author{Chun Yu Weng}
\affiliation{Department of Physics, Hong Kong University of Science and Technology, Clear Water Bay, Hong Kong, China}

\author{Hoi Chun Po}
\affiliation{Department of Physics, Hong Kong University of Science and Technology, Clear Water Bay, Hong Kong, China}
\affiliation{Center for Theoretical Condensed Matter Physics, Hong Kong University of Science and Technology, Clear Water Bay, Hong Kong, China}

\begin{abstract}
Landau levels in certain models are known to protrude into the zero-field energy gap. These are known as anomalous Landau levels (ALLs). We study whether ALLs can lead to Fermi-surface like quantum oscillation in the absence of a zero-field Fermi surface. 
Focusing on two-dimensional multi-band low-energy models of electrons with continuous rotation symmetry, we show that an effective-band description, akin to the semiclassical treatment of Landau level problems in metals, can be used to predict the Landau level spectrum, including possible ALLs. This description then describes ALL induced quantum oscillation for certain insulating models, which we demonstrate through numerical calculations.
\end{abstract}

\maketitle

\section{Introduction}

Physical observables, like conductivity and magnetization, can oscillate periodically as a function of external magnetic field strength. Such phenomena are collectively known as quantum oscillation (QO), and they provide a valuable experimental probe of the Fermi surfaces of electronic materials \cite{Shoenberg1984,Landau30}.
The origin of QO in metals can be attributed to the constant crossing of discrete Landau levels (LL) with the chemical potential, and is generally well-treated by the semiclassical theory pioneered by Lifshitz and Onsager~\cite{Onsager52,LK58,hartnoll10,alexandra18}. For a two-dimensional (2D) spectrum $E(k_{x},k_{y})$, the semiclassical LLs in magnetic field $B$ and the corresponding closed orbits are determined by $\mathcal{S}(E_{n})=2\pi \frac{eB}{\hbar c}(n+\kappa)$, in which $\mathcal{S}(E_{n})$ is the enclosed momentum-space area of the constant energy contour for the $n$th LL energy $E_{n}$ and $\kappa$ takes into account boundary conditions (the Berry phase term). Such formula implies the frequency of the QO is proportional to the area of Fermi surface. 
Conversely, no QO is expected in an insulator or a semimetal with a vanishing Fermi surface.

Contrary to the prediction from the mentioned semiclassical theory, recent experimental discoveries have suggested that QO could also be observed in various insulating systems \cite{zhang23,cooper2015,Wu24}, like $\text{SmB}_{6}$ \cite{Tan15,GLi14,Thomas19}, $\text{YbB}_{12}$ \cite{Xiang18,Mizzi24,Sato19}, $\text{ZrTe}_{5}$ \cite{WangH18}, InAs/GaSb quantum wells \cite{XiaoD19,HanZ19} and Kitaev material $\alpha-\text{RuCl}_{3}$ \cite{czajka21,leeb21}. Different possible mechanisms have been proposed, such as the existence of periodic gap narrowing \cite{Lee21, He21}, surface states \cite{Erten16}, neutral fermi surface \cite{Sodemann18}, and disorder states \cite{shen18}. The active experimental and theoretical investigations on QO in insulators invite one to reconsider if LLs can protrude into the zero-field energy gaps of the system, and, if yes, whether or not they can lead to QO similar to what the semiclassical theory predicts for a non-zero Fermi surface.

LLs can indeed appear at energies at which no zero-field state exists, and these are referred to as 
anomalous Landau levels (ALLs) \cite{xiao03,Yang20,Lian20,devakul21}.  
In fact, as a matter of principle, such phenomenon can be readily anticipated from the Zeeman effect. In the weak-field limit, the semiclassical dispersion of the electron could be modified to $\epsilon(\boldsymbol{k})=\epsilon^{0}(\boldsymbol{k})+\mu(\boldsymbol{k})B$, where $\mu(\boldsymbol{k})$ is the magnetic moment. $\mu(\boldsymbol{k})$ has two origins, the electron spin and the orbital effect. The orbital contribution to $\mu(\boldsymbol{k})$ can be analyzed within a semiclassical framework \cite{Roth66,ChangNiu95,ChangNiu96,ChangNiu08,Koshino10,Gaoniu17,Fuchs18,Atencia24}, throughout this paper we will refer to the orbital contribution as orbital Zeeman term. The existence and properties of ALLs can also be investigated from the lens of quantum geometry \cite{rhim19,Yang20,Yang2021,Jung24,Oh24}.

If the (orbital or spin) Zeeman shift brings the effective dispersion into the original energy gap, then it appears possible for the LLs to go across the original chemical potential and lead to QO.
In practice, however, the applicability of this mechanism is unclear. Typically, the magnitude of the magnetic moment and, associated with that, the Zeeman effect is small. Therefore, the quantum limit has to be reached before the LLs can penetrate substantially into the original band gap, in which case no QO will be observable even if the LLs eventually cross the original Fermi level.
On top of that, the LL quantization, as in the Lifshitz-Onsager theory, suggests that the downward movement of the band bottom (or upward movement of the band top) in a magnetic field does not necessarily imply a similar movement of the LLs.
For instance, consider non-relativistic free electron subjected to the spin Zeeman effect, which is described by the Pauli Hamiltonian $\epsilon(\boldsymbol{k})=\frac{\boldsymbol{k}^{2}}{2m}+\frac{e\hbar}{2mc}B\sigma^{z}$ (henceforth, we take $e=c=\hbar=1$).
Although the spin-down band will move below zero energy, the lowest LL remains at $\epsilon=0$ and there are no ALLs in this case.

In this work, we develop a simple framework aiming at illustrating the presence and origin of ALLs and discussing how they could lead to Fermi-surface like QO in a band insulator systematically. To do so, we consider low-energy Hamiltonians with continuous rotation invariance. While an approximation, such Hamiltonians arise frequently in crystalline materials when the single-particle states closest to the Fermi level sit at high-energy momenta, as is famously the case of Dirac cones in graphene.
Due to the assumed rotation invariance, the exact LL spectrum for such low-energy models can be readily solved by going to the occupation number basis of the associated ladder operators. Here, we leverage such exact solvability and show that an effective band picture can generally be defined, which when supplemented with quantization rules can reproduce the LL sequence similar to that from the semiclassical theory. Different from Kohn's perturbative approach \cite{Kohn59,Blount62,Zak68,Nenciu91}, our effective-band picture generates the exact LLs of the low-energy systems, revealing the evolvement of the LLs by the structure of the effective-Hamiltonians. 
As one might anticipate, the relative angular momenta of the energy bands play a crucial role in controlling the motion of the effective bands and therefore the LLs as a function of $B$ \cite{zhang2005,Neto09,Yang20}.
In particular, the motion of the effective bands in our framework generally depends non-linearly on $B$, and therefore capture corrections beyond the leading-order Zeeman-like term.

Besides rotationally invariant continuum models, we also apply the effective-band analysis to the spin-orbit coupled Lieb lattice model in the weak-field limit. The resulting LLs matches the Hofstadter spectrum well, and the comparison reveals nature of isolated lines and regions in the Hofstadter butterfly.
The effective-band picture is useful in analyzing possible QO resulting from ALLs crossing the chemical potential by sheding a light on its general features and mechanism; such QO is called ALL induced QO. In particular, it is expected that Fermi-surface like QO should result from insulators with small gaps \cite{cooper17} such that there are enough LLs appearing in the gap to form a sequence. Using the effective-band picture we consolidate such expectation by obtaining a formula which computes the number of ALLs that are going to cross the chemical potential with varying magnetic field. 
Besides these, the effective-band picture also helps to answer the question of period of the ALL induced QO and how it is modified from the Lifshitz-Onsager theory.

\section{Low-energy Hamiltonians with rotation symmetry}

Let $\mathcal H_{\boldsymbol{k}}$ be the (low-energy) Hamiltonian of a two-dimensional system with continuous translation and rotational symmetry within the plane. In other words, 
under momentum rotation $\boldsymbol{k}\rightarrow \boldsymbol{k}_{\theta}=(\cos\theta k_{x}+\sin\theta k_{y},\cos\theta k_{y}-\sin\theta k_{x})^{T}$ (or equivalently $k_{-}\rightarrow e^{i\theta}k_{-}$, in which $k_{\pm}=k_{x}\pm ik_{y}$), 
there exists a unitary matrix $\mathcal{U}(\theta)$ such that
\begin{equation}
\label{Uonetransform}
\mathcal{H}_{\boldsymbol{k}_{\theta}}(\theta)=\mathcal{U}(\theta)\mathcal{H}_{\boldsymbol{k}}\mathcal{U}^{\dagger}(\theta),\qquad \mathcal{U}(\theta)=e^{i\theta \hat{\mathcal{Q}}},
\end{equation}
where $\hat{\mathcal{Q}}$, the generator of the $U(1)$ rotation, encodes how the rotation symmetry is represented in the problem. 
For an infinitesimal angle $\theta\rightarrow 0$, Eq. \eqref{Uonetransform} becomes
\begin{equation}
\label{adjointaction}
\mathcal{H}_{\boldsymbol{k}}(\theta)\rightarrow 
\mathcal{H}_{\boldsymbol{k}}+i\theta[\hat{\mathcal{Q}},\mathcal{H}_{\boldsymbol{k}}].
\end{equation}
Matrix $\hat{\mathcal Q}$ contains the information on the relative angular momentum of the states in the low-energy Hilbert space. To fix the ambiguous overall phase of $\mathcal{U}(\theta)$, we require $\hat{\mathcal{Q}}$ to be traceless.
Generally, one may choose a $\boldsymbol{k}$-independent basis in which $\hat{\mathcal Q}$ is diagonal, namely $\hat{\mathcal{Q}}=\text{diag}(\mathcal{Q}_{11},\mathcal{Q}_{22},\cdots)$, we refer to such choice as the canonical basis in the following. 
Importantly, in the canonical basis the diagonal entries of $\hat{\mathcal Q}$ do not need to be integer-valued; rather, the relative angular momenta, which can be reconciled with the differences between any pair of eigenvalues of $\hat{\mathcal Q}$, have to be integer-valued. This can be seen by considering transformation of off-diagonal elements $\mathcal{H}_{ij}$ in the canonical basis, $\mathcal{H}_{ij}(\boldsymbol{k}_{\theta})=\exp{[i(\mathcal{Q}_{ii}-\mathcal{Q}_{jj})\theta]}\mathcal{H}_{ij}(\boldsymbol{k})$, combined with the requirement that the Hamiltonian goes back to itself when $\theta$ is multiples of $2\pi$ (that is, $\boldsymbol{k}_{2\pi}=\boldsymbol{k}$). In particular, the relative angular momenta between neighboring bands are called the indices of the Hamiltonian. In practice the indices are defined by looking specifically at eigenstates at momentum zero $|\boldsymbol{k}=0\rangle$ among all the bands; under rotation of momentum, these states remain unchanged and acquire just extra phases, which define the angular momenta. 
Generally there are $(n-1)$ integer indices for $n$-band models; but for Hamiltonians with decoupled blocks and therefore multiple independent $U(1)$ symmetries, (like $\mathcal{H}^{(+\nu)}\oplus\mathcal{H}^{(-\nu)}$ which we will encounter later), there can be less than $(n-1)$ indices as the relative angular momenta between decoupled blocks are not well-defined.

To relate the form of rotation-invariant canonical Hamiltonians to the indices, we start from a general $n$-band Hamiltonian
$\mathcal{H}_{\boldsymbol{k}}$. 
We can expand $\mathcal{H}_{\boldsymbol{k}}=h_{0}(\boldsymbol{k})\hat{I}+\sum_{i=1}^{n^{2}-1}h_{i}(\boldsymbol{k})\hat{T}_{i}$, in which $\hat{T}_{i}$, $i=1,2,\cdots,(n^{2}-1)$ are the traceless, Hermitian generators of $SU(n)$. The term proportional to the identity is given by $h_{0}(\boldsymbol{k})=\frac{1}{n}\text{tr}\mathcal{H}_{\boldsymbol{k}}$. Rotational symmetry requires that the energy spectrum $E_{k}$ is a function of the modulus of momentum $k=|\boldsymbol{k}|$; so $h_{0}(\boldsymbol{k})$ is a function of $k$. For two-band models, the $SU(2)$ generators are given by the Pauli matrices divided by two. The Hamiltonian is written as $\mathcal{H}_{\boldsymbol{k}}=\hat{D}+(w(\boldsymbol{k})\hat{\sigma}_{+}+\text{h.c.})$, in which $\hat{D}(k)=h_{0}\hat{I}+h_{3}\hat{\sigma}_{z}$ is the diagonal piece. The generator of rotation in the canonical basis is given by $\hat{\mathcal{Q}}=\nu\frac{\hat{\sigma^{z}}}{2}$, characterized by a single integer $\nu$. 
For infinitesimal $\theta$ we have $\mathcal{H}_{\boldsymbol{k}}(\theta)\rightarrow\mathcal{H}_{\boldsymbol{k}}+\frac{i}{2}\nu\theta [\hat{\sigma}^{z},\mathcal{H}_{\boldsymbol{k}}]$, implying that $w(\boldsymbol{k})\rightarrow e^{i\theta \nu}w(\boldsymbol{k})$ for finite $\theta$. So the canonical two-band models are given by, to the leading order in the off-diagonal terms, 
\begin{equation}
\label{Hk1}
\mathcal{H}_{\boldsymbol{k}}^{(\nu)}=\left(\begin{array}{cc}
\cdots&\gamma k_{-}^{\nu}\\
\gamma k_{+}^{\nu}&\cdots
\end{array}\right).
\end{equation}
Here and hereafter the negative powers of momentum are always understood as $(k_{\mp})^{-\nu}\rightarrow (k_{\pm})^{+\nu}$. For three-band models, similar analysis \cite{SM} shows that the generator in the canonical basis reads $\hat{\mathcal{Q}}=\text{diag}(\frac{2p}{3}+\frac{q}{3}, -\frac{p}{3}+\frac{q}{3},-\frac{p}{3}-\frac{2q}{3})$ with two integers $(p,q)$ characterizing the relative angular momenta. The canonical Hamiltonian is given by
\begin{equation}
\label{threebandpq}
\mathcal{H}_{\boldsymbol{k}}^{(p,q)}=\left(\begin{array}{ccc}
\cdots&\gamma_{1}(k_{-})^{p}&\gamma_{2}(k_{-})^{p+q}\\
\gamma_{1}(k_{+})^{p}&\cdots&\gamma_{3}(k_{-})^{q}\\
\gamma_{2}(k_{+})^{p+q}&\gamma_{3}(k_{+})^{q}&\cdots
\end{array}\right).
\end{equation}
In both (\ref{Hk1}) and (\ref{threebandpq}), the $\cdots$ denotes diagonal terms which are rotationally invariant (quadratic in $k$ or higher); for example, they can take the form $\alpha (k^{4}-k_{0}^{4})+\beta(k^{2}-k_{0}^{2})$ which crosses energy zero at a ring of momentum $k=k_{0}$. For parabolic bands the parameters $\beta>0$ has the dimension of mass-inverse. The approach can be readily generalized to higher number of bands, in general for $n$-band model in canonical basis, the off-diagonal elements of the Hamiltonian reads $\mathcal{H}_{ij}=\gamma_{ij}(k_{-})^{\mathcal{Q}_{ii}-\mathcal{Q}_{jj}}$.

\section{Landau levels in the occupation number basis and extra ansatz states}

Next we sketch the standard procedure for obtaining LLs in these low-energy models. Inside a magnetic field $B$, one introduces the Peierls substitution $\boldsymbol{k}\rightarrow \tilde{\boldsymbol{k}}=\boldsymbol{k}-\boldsymbol{A}$. Because of the commutator $[\tilde{k}_{x},\tilde{k}_{y}]=iB$, we introduce the bosonic ladder operator $\tilde{k}_{-}=\tilde{k}_{x}-i\tilde{k}_{y}=\sqrt{2B}a^{\dagger}$, and $\tilde{k}_{x}^{2}+\tilde{k}_{y}^{2}=2B(a^{\dagger}a+\frac{1}{2})$. Among different terms, the products of $k_{\pm}$ and $k^{2}$, such as $k^{q}(k_{\pm})^{p}$, are anomalous because such terms cannot be translated into $a$ and $a^{\dagger}$ operators unambiguously. For example, $k^{2}k_{+}=k_{+}k^{2}$ without magnetic field, but they are not equal when translated into $a$ and $a^{\dagger}$. So for low-energy Hamiltonians such as (\ref{Hk1}) and (\ref{threebandpq}), we enforce the following consistency requirement: if a off-diagonal term contains nonzero power of $k_{\pm}$, the coefficient $\gamma$ is constant; otherwise, $\gamma$ is a function of $k^{2}$. With ladder operators, the Hamiltonians \eqref{Hk1} and \eqref{threebandpq} inside magnetic field can be readily obtained \cite{SM}. Using the bosonic states $|n\rangle$ one can solve the canonical two-band Hamiltonian $\mathcal{H}_{B}^{(\nu)}$ by ansatz solutions: $(|n\rangle, |n-\nu\rangle)^{T}$, with the series of $n$ starting from $\text{min}\{n,n-\nu\}=0$, and obtain the LL sequence $E_{n}$. For three-band models, the general ansatz solution to the canonical Hamiltonian $\mathcal{H}_{B}^{(p,q)}$ is given by $(|n\rangle, |n-p\rangle, |n-p-q\rangle)^{T}$, with $\text{min}\{n,n-p,n-p-q\}=0$ defining the series of $n$. Since the rotation of momentum becomes a phase of the bosonic operator $a^{\dagger}\rightarrow e^{i\theta}a^{\dagger}$ inside magnetic field, the difference of band angular momenta are reflected in the boson occupation numbers. 

For models with non-zero indices, there are extra ansatz states (including possible null states) that are not captured by the LL sequence, for example $(|0\rangle,0)^{T}$ and $(|1\rangle,0)^{T}$ for $\mathcal{H}_{B}^{(2)}$. In general for two-band models, the number of extra ansatz states equals to $|\nu|$ while the number of extra ansatz states in three-band models is given by $\text{max}\{|p|,|q|,|p+q|\}$. Note that this is not the number of extra LLs since each ansatz state often corresponds to multiple LL solutions \cite{SM}. The behavior of extra ansatz states with varying magnetic field is distinct from the LL sequence and can be observed in both the effective bands and the Hofstadter butterfly for tight-binding models (see examples below). The appearance of extra ansatz solutions is another indicator for relative angular momentum between bands.
To reduce cluttering, in the following we refer to the series of solution labeled by integer $n$ as the LL sequence. The whole set of LLs, comprising both the LL sequence and the extra ansatz solutions, is referred to as the LL spectrum.

While correct, the number basis calculation does not readily reveal the behavior of the system, especially on how ALLs can emerge. 
It is also unclear how the number basis calculation can be reconciled with a semiclassical treatment which is usually sufficient in the weak-field limit.
For instance, consider a parabolic band with non-zero orbital magnetic moment. From Chang-Niu semi-classical wave-packet approach \cite{ChangNiu95,ChangNiu96}, one can deduce the existence of an orbital Zeeman shift which captures the corrections to the Onsager quantization and could lead to ALLs \cite{Yang2021}. 
It is unclear how the expected orbital Zeeman shift arises from the number-basis calculation even when the latter is applicable.
In the following, we propose an effective band description which can bridge the two approaches and provide a framework for analyzing the emergence of ALLs and ALL induced QO in band insulators.

\section{The effective-band picture}

\begin{figure}
    \centering
    \includegraphics[width=0.9\linewidth]{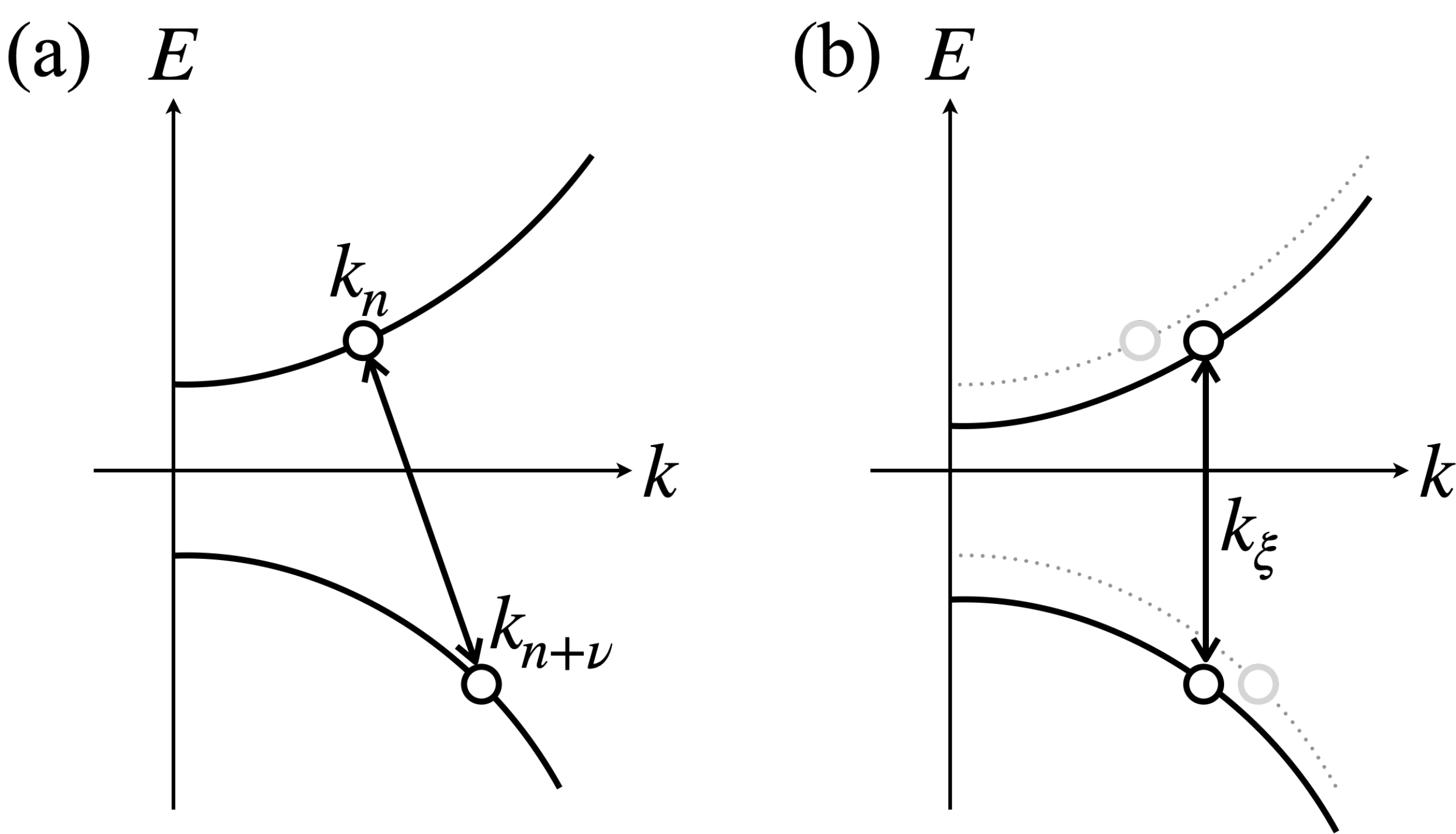}
    \caption{\textbf{Constructing the effective band} (a) In the original model, LLs come from states with (discrete) momentum $k_{n}$ (corresponding to number-basis state $|n\rangle$) hybridizing with $k_{n+\nu}$ due to the relative angular momentum. The curves represent the diagonal terms of the Hamiltonian and the arrow represents the ``hybridization" caused by the off-diagonal terms. (b) The effective band is constructed such that states with $k_{\xi}$ hybridizing with $k_{\xi}$ reproduces the same LLs. Compared to (a), the diagonal terms of the effective bands are shifted downward such that the hybridizing states are now labeled by the same $k_\xi$.
    }
    \label{Fig:cartoon}
\end{figure}

We now introduce the effective-band description.
We start by introducing a continuous parameter
\begin{equation}
\label{orbitalkn}
k_{\xi}=\sqrt{2B \left(\xi+\frac{1}{2}\right)},
\end{equation}
which has the unit of a momentum and $\xi$ is any real number larger than $-1/2$. Although the specific choice of $1/2$ in the definition of $k_\xi$ is immaterial, it simplifies the expressions we obtain below. 
In the original Hamiltonians (\ref{Hk1}) and (\ref{threebandpq}), only states at the same momentum can hybridize with each other by the off-diagonal elements due to the assumed translation invariance. Inside a magnetic field, states with number-basis label $n$ generally hybridizes with states at another label $n'$ because of the relative angular momentum of the bands. We want to construct a new effective Hamiltonian $\mathcal{H}_{E}(k, B)$ which depends on the modulus of momentum $k$ and the strength of the magnetic field $B$, such that its matrix elements evaluated at a chosen series of momenta are identical to the matrix elements of the original number-basis magnetic Hamiltonian under its ansatz solution. For such effective Hamiltonian, the energy eigenvalues at the selection of $k_{\xi}$, combined with the extra ansatz states reproduce the exact LL sequence. The admissible values of $k_\xi$ are determined by a set of accompanying quantization rules. The construction of the effective Hamiltonian can be understood as a simple shift of variables such that states labeled by the same momentum-like parameter $k_\xi$ in \eqref{orbitalkn} are now coupled by the off-diagonal elements of the Hamiltonian, as illustrated in Fig. \ref{Fig:cartoon}. 
The origin and the steps for constructing the effective bands and quantization rules from the original Hamiltonian $\mathcal{H}_{\boldsymbol{k}}$ and the $U(1)$ generator $\hat{\mathcal{Q}}$ are detailed as follows.

We start by bringing the Hamiltonian to the canonical basis in which $\hat{\mathcal{Q}}$ is diagonal. 
Without loss of generality, we assume the diagonal elements of the $m$-band canonical generator $\hat{\mathcal{Q}}$ are $\mathcal{Q}_{ii}$ ($i=1,2,\cdots, m$) with $\mathcal{Q}_{11}\geq \mathcal{Q}_{22}\geq \cdots \mathcal{Q}_{mm}$. These elements satisfy $\sum_{i=1}^{m}\mathcal{Q}_{ii}=0$ and $\mathcal{Q}_{mm}=\zeta<0$ is the most negative element. We define $q_{i}=\mathcal{Q}_{ii}-\zeta\geq 0$ as the relative angular momentum with respect to the $m$th band. When solving the LLs with number-basis calculation, the ansatz solution for canonical Hamiltonian is $(|n+q_{1}\rangle, \cdots, |n+q_{i}\rangle, \cdots, |n\rangle)^{T}$ with $n=0,1,2,\cdots$. To ensure that the effective Hamiltonian have the same matrix elements at chosen momenta with the original Hamiltonian under its specific ansatz states, we have to adjust the elements of the effective Hamiltonian. In particular for the $i$th diagonal element, the momentum squared $\tilde{\boldsymbol{k}}^{2}$ in operator form produces the following eigenvalue under ansatz state,
\begin{eqnarray}
\label{translation}
    \begin{aligned}
        &2B(a^{\dagger}a+\frac{1}{2})|n+q_{i}\rangle \rightarrow \\&2B(n+q_{i}+\frac{1}{2})=2B(n+|\zeta|+\frac{1}{2})+2B\mathcal{Q}_{ii}.
    \end{aligned}
\end{eqnarray}
The translation \eqref{translation} suggests that the operator form of $\tilde{\boldsymbol{k}}^{2}$ acting on the ansatz states $|n+q_{i}\rangle$ in number-basis calculation produces an eigenvalue identical to polynomial $k_{\xi}^{2}+2B\mathcal{Q}_{ii}$ evaluated at $\xi=n+|\zeta|$ with $k_{\xi}$ defined in \eqref{orbitalkn}. The diagonal elements of the effective Hamiltonian can be readily obtained. Generally, the $i$-th diagonal entry of the canonical Hamiltonian contains terms in powers of $k^2$, say $\beta_i k^2 + c_i$.
In the effective band picture, we replace $k^2 \mapsto k^2 + 2 B \mathcal Q_{ii}$ in the $i$-th diagonal entry of the Hamiltonian. When $\mathcal{Q}_{ii} \neq 0$, this replacement is reminiscent of the orbital Zeeman term contributed by an orbital magnetic moment\cite{ChangNiu95,ChangNiu96}. 
The procedure goes through unaltered even when higher powers of $k$ are present in the diagonal term of the Hamiltonian, for example $\alpha_i k^4 + \beta_i k^2 + c_i $ should be replaced by $\alpha_i (k^2 + 2 B \mathcal Q_{ii})^2 + \beta_i (k^2 + 2 B \mathcal Q_{ii}) + c_i $, which then leads to higher order shifts proportional to $B^2$.

The translation \eqref{translation} also determines the quantization rule: the selected momenta are given by $k = k_\xi$ as defined in Eq.\ \eqref{orbitalkn} with $\xi=|\zeta|, 1+|\zeta|, 2+|\zeta|, \cdots$. In other words, the chosen momentum sequence is $k_{n}=\sqrt{2B(n+|\zeta|+\frac{1}{2})}$, with $n=0,1,2,\cdots$.

Next we look at the off-diagonal terms. They generally read $(\mathcal{H}_{\boldsymbol{k}})_{ij}=\gamma_{ij} g_{ij}(k)$ in which $\gamma_{ij}$ are constants and $g_{ij}(k)$ are either powers of $k_{\pm}$ or functions of $k^{2}$. Assuming $\mathcal{Q}_{ii}\geq\mathcal{Q}_{jj}$ the element reads $\mathcal{H}_{ij}=\gamma_{ij}(\sqrt{2B} a^{\dagger})^{q_{i}-q_{j}}$ in the number-basis calculation. If $q_{i}\neq q_{j}$, the off-diagonal element acting on the ansatz solution is denoted by $\gamma_{ij}(\sqrt{2B}a^{\dagger})^{q_{i}-q_{j}}|n+q_{j}\rangle$, which produces the matrix element 
\begin{equation}
\label{offdiagonalelement}
\gamma_{ij}(\sqrt{2B})^{q_{i}-q_{j}}\sqrt{n+q_{j}+1}\sqrt{n+q_{j}+2}\cdots\sqrt{n+q_{i}}
\end{equation}
in the Hamiltonian.
We can express the variable $n$ with the $n$th member of our quantization rule $(k_{\xi})_{n}=\sqrt{2B(|\zeta|+n+\frac{1}{2})}$ in the matrix element \eqref{offdiagonalelement}. This procedure leads to the following results for off-diagonal elements.
Given that $\mathcal{Q}_{ii}\geq \mathcal{Q}_{jj}\geq \zeta$, we define two non-negative numbers $r_{ij}=\mathcal{Q}_{ii}-\mathcal{Q}_{jj}$ and $s_{ij}=\mathcal{Q}_{jj}-\zeta(=q_{j})$. For the case $r_{ij}\geq 1$, $\mathcal{Q}_{ii}\neq \mathcal{Q}_{jj}$, $g_{ij}(k)$ must contain nonzero order of $k_{\pm}$, then the off-diagonal elements of the effective band are given by
\begin{equation}
\label{offdiagonal}
(\mathcal{H}_{E})_{ij}=\gamma_{ij}\prod_{t=1}^{r_{ij}}\sqrt{k^{2}+2B \left(s_{ij}+t-|\zeta|-\frac{1}{2} \right)}.
\end{equation} 
For the case $r_{ij}=0$ on the other hand, $\mathcal{Q}_{ii}=\mathcal{Q}_{jj}$, $g(k)$ is a function of $k^{2}$ and does not contain $k_{\pm}$, then the off-diagonal elements are given by 
\begin{equation}
\label{offdiagonaltwo}
    (\mathcal{H}_{E})_{ij}=\gamma_{ij}g_{ij}(k^{2}+2B\mathcal{Q}_{ii}).
\end{equation}
The construction for effective Hamiltonian $\mathcal{H}_{E}(k,B)$ and the quantization rule is thus completed.

By design, the effective Hamiltonian $\mathcal H_E(k, B)$ has a $B$-dependent spectrum similar to the result in a semiclassical treatment. In simple cases, the diagonal $B$-dependence can be reconciled precisely with an orbital Zeeman shift. However, the general dependence is not linear in $B$, and the modification to the off-diagonal terms cannot be readily interpreted in the semiclassical treatment. Furthermore, together with the stated quantization rule the original LL spectrum is reproduced for any magnetic field, and so the description is valid from the weak-field to the large-field limit as long as the low-energy effective model remains a good approximation.

We note two points of caution before moving on to illustrating the formulation through concrete examples. First, as mentioned before, when there are non-zero relative angular momenta between the states, the extra ansatz LLs are not covered by the mentioned procedure of diagonalizing $\mathcal H_E$ and choosing the allowed energy levels according to the quantization rule, and so have to be computed independently following the standard approach. Second, there is an inherent ambiguity in the definition of $k_\xi$ in Eq.\ \eqref{orbitalkn}, which also affects how $\mathcal H_{E}$ is defined. As mentioned, the choice of $1/2$ in Eq.\ \eqref{orbitalkn} appears arbitrary, and one can indeed replace that by $1/2 + \rho$. However, in doing so the charge operator $\hat {\mathcal Q}$ entering into the definition of $\mathcal H_E$ should also be replaced by $\hat {\mathcal Q} \mapsto  \hat {\mathcal Q}  + \rho \hat \openone$. Similarly, the extra shift of $\rho$ should also be reflected in Eq.\ \eqref{offdiagonal}. Our present choice of $\rho = 0$ is fixed by requiring $\hat {\mathcal Q}$ to be traceless. 

\section{Examples of effective-band picture}

Here we consider application of the effective-band picture to three concrete examples with increasing complexity, starting from simple two-band model and three-band model in canonical form. For these simple models, one observes how the effective bands move and evolve inside magnetic field and how ALLs are formed. We then move on to a tight-binding model with isolated flat band, namely the Lieb lattice model with spin-orbit coupling (SOC). Inside magnetic field, the flat band acquires a finite bandwidth and ALLs are formed. We relate such phenomenon to the arrangement of angular momentum. In particular, the low-energy LLs from effective-band calculation are compared with the Hofstadter spectrum, resulting in good agreement, which further clarifies the nature and origin of branches in Hofstadter butterfly. 

\begin{figure*}[t]
\centering
\includegraphics[width=\textwidth]{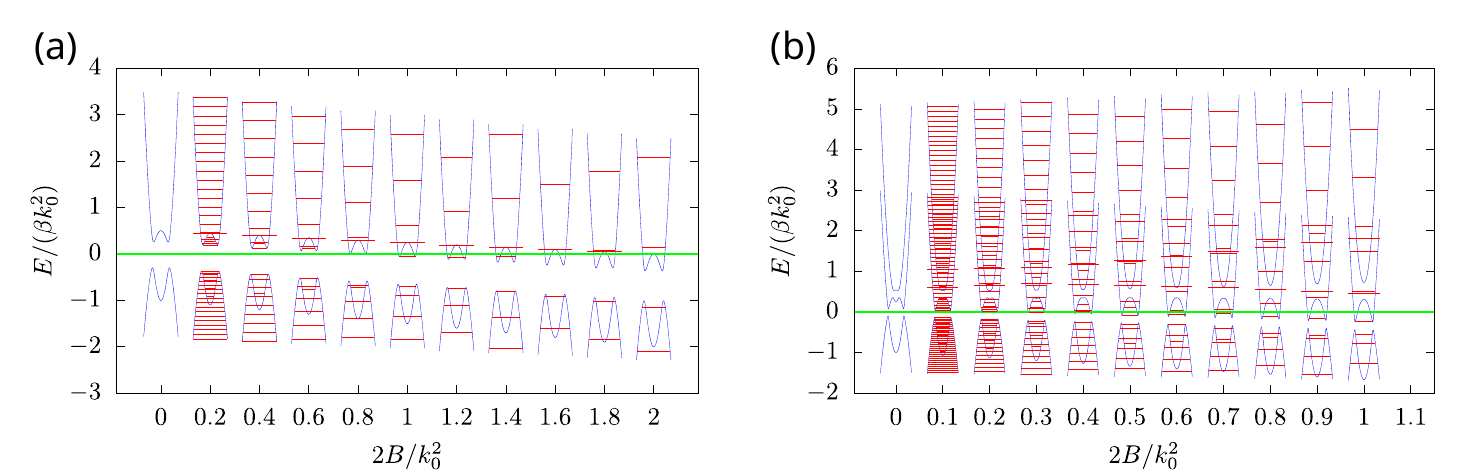}
\caption{\textbf{Examples of effective bands.} (a) The two-band model $\mathcal{H}_{E}^{(-1)}$ with $\frac{\beta'}{\beta}=0.5$, $\frac{\gamma}{\beta k_{0}}=0.3$. (b) The three-band model $\mathcal{H}_{E}^{(1,-1)}$ with $\frac{\beta_{1}}{\beta}=1.2$, $\frac{\beta_{2}}{\beta}=0.5$, $\frac{\gamma_{1}}{\beta k_{0}}=\frac{\gamma_{2}}{\beta k_{0}}=\frac{\gamma_{3}}{\beta k_{0}}=0.1$, $\frac{k_{1}}{k_{0}}=0.5$. LLs are represented by the horizontal line segments. An imaginary chemical potential is set at $\frac{E}{\beta k_{0}^{2}}=0$. The left-most part with $\frac{2B}{k_{0}^{2}}=0$ represents the original bands without magnetic field. The extra solutions are included in the spectrum with slightly longer line segments.}
\label{Fig:EB}
\end{figure*}

Firstly for the canonical two-band $\mathcal{H}_{\boldsymbol{k}}^{(\pm 1)}$ with diagonal piece being parabolic, the generator $\hat{\mathcal{Q}}=\pm\frac{\sigma^{z}}{2}$, the only off-diagonal term $\mathcal{H}_{12}$ has $r_{12}=1$ and $s_{12}=0$, so the effective Hamiltonian is 
\begin{equation}
\label{HEone}
\mathcal{H}_{E}^{(\pm 1)}=\left(\begin{array}{cc}
\beta(k^{2}-k_{0}^{2})\pm B\beta&\gamma k\\
\gamma k&-\beta'(k^{2}-k_{0}^{2})\pm B\beta'
\end{array}\right).
\end{equation}
More generally, the diagonal piece of $\mathcal{H}^{(\pm 1)}$ can be taken as higher-order terms (quartic in $k$), namely, $\mathcal{H}^{(\pm 1)}_{11}=\alpha(k^{4}-k_{0}^{4})+\beta(k^{2}-k_{0}^{2})$ and $\mathcal{H}^{(\pm 1)}_{22}=-\alpha'(k^{4}-k_{0}^{4})-\beta'(k^{2}-k_{0}^{2}))$. Without off-diagonal terms, the two diagonal terms cross at $|\boldsymbol{k}|=k_{0}$ with energy zero. In an external magnetic field $B$, the diagonal piece of the effective Hamiltonian \eqref{HEone} changes to $(\mathcal{H}_{E}^{(\pm 1}))_{11}=\alpha[(k^{2}\pm B)^{2}-k_{0}^{4}]+\beta(k^{2}\pm B-k_{0}^{2})$ and $(\mathcal{H}_{E}^{(\pm 1)})_{22}=-\alpha'[(k^{2}\mp B)^{2}-k_{0}^{4}]-\beta'(k^{2}\mp B-k_{0}^{2})$. The quantization rule is $\xi=\frac{1}{2}, 1+\frac{1}{2},2+\frac{1}{2},\cdots$, so the selected momenta for LLs are $k_{\xi}=\sqrt{2Bn}$ for $n=1,2,3,\cdots$. Beside these, $\mathcal{H}^{(+1)}$ has one extra ansatz state that is not captured in the sequence, $(|0\rangle,0)^{T}$, with energy $E_{\text{ex}}=\alpha(B^{2}-k_{0}^{4})+\beta(B-k_{0}^{2})$. Similarly, $\mathcal{H}^{(-1)}$ has one extra state $(0,|0\rangle)^{T}$, with energy $E_{\text{ex}}=-\alpha'(B^{2}-k_{0}^{4})-\beta'(B-k_{0}^{2})$.
In Fig. \ref{Fig:EB}, we plot the effective bands as a function of $B$, which illustrates their evolution. As shown in the Fig. \ref{Fig:EB} (a), the movement of the effective band $\mathcal{H}_{E}^{(-1)}$ corresponds to a simple orbital Zeeman shift, the LLs carried into the original band gap form ALLs, some of which cross the (imaginary) chemical potential set at zero. As expected, the direction of effective-band movement reverses for opposite sign of $\nu$ (the gap moves upwards for $\mathcal{H}_{E}^{(+1)}$).

Secondly for three-band canonical $\mathcal{H}_{\boldsymbol{k}}^{(1,-1)}$, we take the diagonal terms in (\ref{threebandpq}) to be parabolic,
\begin{equation}
\label{threebm}
\mathcal{H}_{\boldsymbol{k}}^{(1,-1)}=\left(\begin{array}{ccc}
\beta_{1}(k^{2}+k_{1}^{2})&\gamma_{1}k_{-}&\gamma_{2}k_{0}\\
\gamma_{1}k_{+}&\beta(k^{2}-k_{0}^{2})&\gamma_{3}k_{+}\\
\gamma_{2}k_{0}&\gamma_{3}k_{-}&-\beta_{2}(k^{2}-k_{0}^{2})
\end{array}\right).
\end{equation}
As $\hat{\mathcal{Q}}=\text{diag}(\frac{1}{3},-\frac{2}{3},\frac{1}{3})$, the quantization rule is $\xi=\frac{2}{3},1+\frac{2}{3},2+\frac{2}{3},\cdots$. The off-diagonal terms $(\mathcal{H}_{\boldsymbol{k}})_{12}$ and $(\mathcal{H}_{\boldsymbol{k}})_{23}$ have $r_{12}=r_{23}=1$ and $s_{12}=s_{23}=0$, while $(\mathcal{H}_{\boldsymbol{k}})_{13}$ has $r_{13}=s_{13}=0$. Following the receipe, the effective Hamiltonian is given by
\begin{widetext}
\begin{equation}
\label{threebsb}
\mathcal{H}_{E}^{(1,-1)}=\left(\begin{array}{ccc}
\beta_{1}(k^{2}+k_{1}^{2})+\beta_{1}(2B)(\frac{1}{3})&\gamma_{1}\sqrt{k^{2}-\frac{B}{3}}&\gamma_{2}k_{0}\\
\gamma_{1}\sqrt{k^{2}-\frac{B}{3}}&\beta(k^{2}-k_{0}^{2})+\beta(2B)(-\frac{2}{3})&\gamma_{3}\sqrt{k^{2}-\frac{B}{3}}\\
\gamma_{2}k_{0}&\gamma_{3}\sqrt{k^{2}-\frac{B}{3}}&-\beta_{2}(k^{2}-k_{0}^{2})-\beta_{2}(2B)(\frac{1}{3})
\end{array}\right)
\end{equation}
\end{widetext}
There is one extra ansatz state $(|0\rangle, 0,|0\rangle)^{T}$, correspondingly there are two extra LLs \cite{SM}. As shown in Fig. \ref{Fig:EB} (b), the effects of $B$ field goes beyond a simple orbital Zeeman term with modification of off-diagonal terms at the same time. Consequently the effective bands change shape while they evolve with $B$ field. ALLs are observed which cross the (imaginary) chemical potential at zero.

We next discuss a more realistic tight-binding model, which was also studied in Ref.\ \onlinecite{Yang2021} from a different perspective. Unlike the two simple models above which only include dispersive bands, flat-band systems \cite{guo09,Weeks10} represent another natural setup for the emergence of ALLs as the LLs may not remain exactly degenerate at the original flat-band energy and instead develop a finite bandwidth and enter the original energy gap\cite{Yang2021} .
To illustrate how such phenonomona can be understood in the effective-band picture, we consdier the Lieb lattice model with spin-orbit coupling (SOC) \cite{Weeks10}, which has an isolated flat band \cite{Green10}. (Similar discussion can be applied to systems with flat-band touching with a dispersive band, see Supplemental Materials \cite{SM} for details.)

\begin{figure*}[t]
\centering
\includegraphics[width=\textwidth]{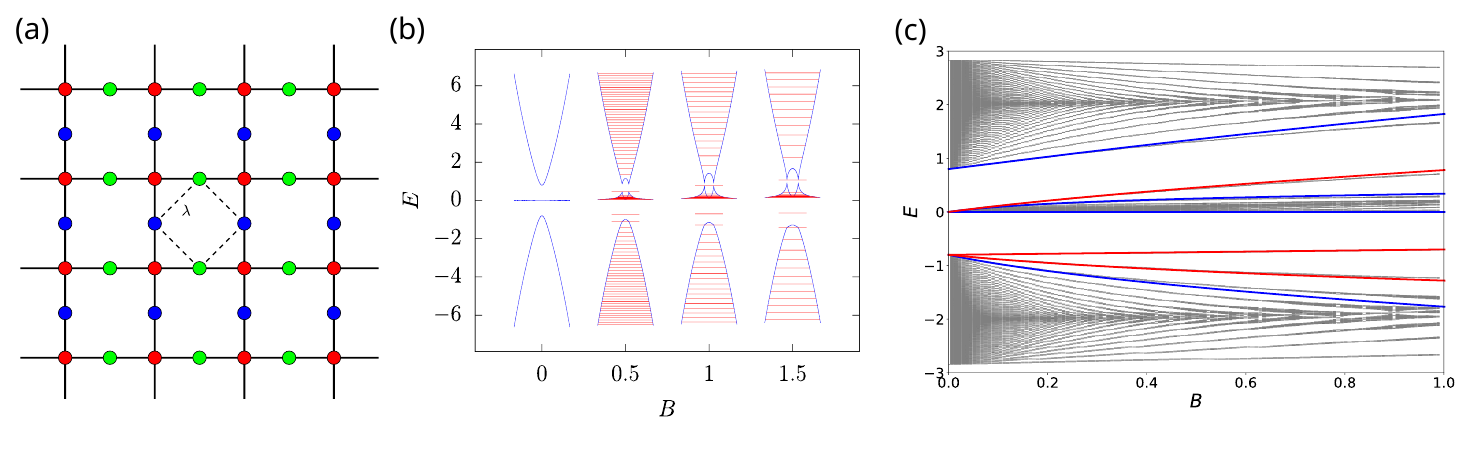}
\caption{\textbf{ALLs in the Lieb lattice model}.
(a) The Lieb lattice model with SOC. (b) Effective band of the Lieb lattice model with $\lambda=0.2$ as a function of magnetic field, the extra LLs are shown with slightly longer line segments. (c) Hofstadter butterfly of the Lieb lattice model in the small field region, with $\lambda=0.2$. The Hofstadter spectrum is compared with ``special" LLs computed from the effective-band picture with the same $\lambda$. The upper and lower blue lines are the bottom (top) LL of the upper (lower) band; red lines are the three extra LLs; the region between the blue lines in the middle is the flat band that gets broader.}
\label{Fig:lieb}
\end{figure*}

On the Lieb lattice (Fig. \ref{Fig:lieb} (a)) we start with an electron-hopping model between nearest neighbors $\mathcal{H}=\sum_{\langle ij \rangle,\sigma}c_{i\sigma}^{\dagger}c_{j\sigma}+\text{h.c.}$, the hopping amplitude is taken to be one for simplicity \cite{Weeks10,Yang2021}. We then add SOC terms for next-nearest-neighbors $\mathcal{H}_{\rm SOC}=i\lambda\sum_{\langle \langle ij\rangle \rangle}\nu_{ij}c_{i\alpha}^{\dagger}(\hat{\sigma}_{z})_{\alpha\beta}c_{j\beta}$, with $\nu_{ij}=\pm 1$. The spin up and down electrons remain decoupled with SOC, for one spin component (say up), the Hamiltonian in momentum space is given by \cite{Weeks10}
\begin{equation}
\mathcal{H}(\tilde{\boldsymbol{k}})=\left(\begin{array}{ccc}
0&2\cos\frac{\tilde{k}_{y}}{2}&-4i\lambda\sin\frac{\tilde{k}_{x}}{2}\sin\frac{\tilde{k}_{y}}{2}\\
2\cos\frac{\tilde{k}_{y}}{2}&0&2\cos \frac{\tilde{k}_{x}}{2}\\
4i\lambda\sin\frac{\tilde{k}_{x}}{2}\sin\frac{\tilde{k}_{y}}{2}&2\cos\frac{\tilde{k}_{x}}{2}&0
\end{array}\right).
\end{equation}
Its band structure is formed by one flat band $E=0$ and two dispersive bands $E=\pm 2\sqrt{\cos^{2}\frac{\tilde{k}_{x}}{2}+\cos^{2}\frac{\tilde{k}_{y}}{2}+4\lambda^{2}\sin^{2}\frac{\tilde{k}_{x}}{2}\sin^{2}\frac{\tilde{k}_{y}}{2}}$.
Without SOC, the three bands in the model are degenerate at $(\pi, \pi)$, which are then gapped out for $\lambda \neq 0$.
For small $|\lambda|$, the low-energy Hamiltonian is described by states around $\tilde{\boldsymbol{k}} = (\pi,\pi)$, and so we expand the Hamiltonian around this point by defining $k_{x,y} = \tilde k_{x,y}-\pi$.
Keeping to the quadratic order, the low-energy Hamitonian is given by
\begin{equation}\label{eq:lieb_low_energy}
\mathcal{H}(\boldsymbol{k})=\left(\begin{array}{ccc}
0&-k_{y}&- i \frac{1}{2} \lambda(2-k_{x}^{2} - k_{y}^{2})\\
-k_{y}&0&-k_{x}\\
i \frac{1}{2} \lambda(2-k_{x}^{2} - k_{y}^{2})&-k_{x}&0
\end{array}\right).
\end{equation}
Written out in the sublattice basis, Eq.\ \eqref{eq:lieb_low_energy} is not in the canonical form. By a basis rotation, $\mathcal{H}(\boldsymbol{k})\rightarrow \tilde{\mathcal{H}}(\boldsymbol{k})=U\mathcal{H}(\boldsymbol{k})U^{\dagger}$ with
\begin{equation}
U=\frac{1}{\sqrt{2}}\left(\begin{array}{ccc}
1&0&-i\\
0&\sqrt{2}&0\\
1&0&i
\end{array}\right),
\end{equation} 
the Hamiltonian becomes
\begin{equation}
\label{Hthreeone}
\tilde{\mathcal{H}}(\boldsymbol{k})=\frac{1}{2}\left(\begin{array}{ccc}
  \lambda(8-k^{2})&\sqrt{2}ik_{+}&0\\
   -\sqrt{2}ik_{-}&0&\sqrt{2}ik_{+}\\
   0&-\sqrt{2}ik_{-}&-\lambda(8-k^{2})
\end{array}
\right).
\end{equation}
We further note the relation
\begin{equation}\label{eq:chiral_like}
    \Gamma_{\boldsymbol{k}} \tilde{\mathcal{H}}(\boldsymbol{k}) \Gamma_{\boldsymbol{k}} = - \tilde{\mathcal{H}}(\boldsymbol{k}),\quad
    \Gamma_{\boldsymbol{k}} = 
    \left(\begin{array}{ccc}
    0 & 0 & e^{-i 2 \theta}\\
    0 & 1 & 0 \\
    e^{i 2 \theta} & 0 & 0
    \end{array}
    \right),
\end{equation}
where $\Gamma_{\boldsymbol{k}}^\dagger = \Gamma_{\boldsymbol{k}}$ and $\theta $ is defined through $k_\pm = k e^{\mp i \theta}$. $\Gamma_{\boldsymbol{k}}$ acts like a chiral symmetry and ensures the spectrum of $\tilde{\mathcal{H}}(\boldsymbol{k})$ is particle-hole symmetric \cite{Chiu16}. Notice that Eq.\ \eqref{Hthreeone} has rotational symmetry with generator $\hat{\mathcal{Q}}=\text{diag}(-1,0,1)$. Its associated effective Hamiltonian $\mathcal{H}_{E}^{(-1,-1)}$ is
\begin{equation}
\label{effectH}
\frac{1}{2}\left(\begin{array}{ccc}
\lambda[8-(k^{2}-2B)]&i\sqrt{2(k^{2}-B)}&0\\
-i\sqrt{2(k^{2}-B)}&0&i\sqrt{2(k^{2}+B)}\\
0&-i\sqrt{2(k^{2}+B)}&\lambda[(k^{2}+2B)-8]
\end{array}\right),
\end{equation}
and the quantization rule is $\xi=1,2,3,\cdots$. This model has two extra ansatz states and three extra LLs \cite{SM}. 

For the evolving effective bands in magnetic field, what was originally the middle flat band first gets broader then mixes with one of the dispersive bands, as shown in Fig. \ref{Fig:lieb} (b). Before the mixing, the middle band has finite bandwidth but hosts infinite number of LLs. The bandwidth can be taken as a rough estimate of ``spread of LLs" \cite{Yang20}. 
Curiously, the broadening of the original zero-energy flat band in the magnetic field can be attributed to the arrangement of the relative angular momenta in the present problem. 
From Eq.\ \eqref{effectH}, one can observe that the chiral-symmetry-like relation in Eq.\ \eqref{eq:chiral_like} is broken for any $B\neq 0$, which then allows for a broadening of the zero-energy flat band in the effective band picture.
In contrast, the relation in Eq.\ \eqref{eq:chiral_like} will be preserved for finite $B$ for the case of three-band model $\mathcal{H}^{(\nu,-\nu)}$, which differs from Hamiltonian \eqref{Hthreeone} only in the arrangement of $k_{\pm}$ of off-diagonal elements. Consequently the LLs of the flat band will remain pinned at zero energy for $\mathcal{H}^{(\nu,-\nu)}$.

To illustrate our results, we also compare the effective bands with the the Hofstadter spectrum of the original tight-binding model. As can be seen in Fig. \ref{Fig:lieb} (c), the edges of the Hofstadter spectrum in the small field limit match well with that predicted from the effective band description.
Such comparison also helps clarify the origin and nature of the different branches and isolated lines in the Hofstadter butterfly \cite{Taut2005}. In particular, we point out two observations. Firstly, the ALLs, being LLs that evolve into the gap of zero-field limit, are clearly observable in the Hofstadter butterfly. Secondly, the extra ansatz states, resulting from non-zero relative angular momenta, can show up as isolated lines in the small-field region of the Hofstadter butterfly.

\section{Effective-band picture and anomalous Landau level induced quantum oscillation}

Having illustrated the effective band picture through various examples, we next turn to the question of how the effective-band picture could help analyze the ALL induced Fermi-surface like quantum oscillations.
There are two basic requirements: first, we require the existence of ALLs which cut across the original Fermi level; second, we require these ALLs to be part of a LL sequence resembling that from the Lifshitz-Onsager semi-classical quantization of a Fermi surface in the weak-field limit. These points are perhaps best illustrated by non-examples. For instance, in the SOC Lieb lattice there are extra ansatz LLs which can even traverse the whole energy gap defined in the zero-field limit. However, these isolated ALLs do not lead to quantum oscillations in physical observables. Alternatively, if the LL sequence reaches the Fermi level only in the quantum limit, there will not be enough number of LL crossings to provide a well-defined oscillation frequency. In the following, we discuss how the effective-band picture provides a way to identify insulating models which satisfy the two conditions laid out above and therefore feature ALL induced Fermi-surface like quantum oscillation. Furthermore, we use the effective-band picture to analyze the features of ALL induced QO.

Consider a band insulator with nonzero indices, such that the effective bands could evolve as a function of the magnetic field $B$.
With the band movement, the effective-band gap can be carried away from the chemical potential, as shown in Fig. \ref{Fig:EB}. In this case, there is a critical magnetic field $B_{c}$ under which the chemical potential starts crossing the effective band. On the other hand, the model can be reduced to multiple independent components which share the same energy gap, and the gaps move in opposite directions under magnetic field. For example, the four-band model $\mathcal{H}=\mathcal{H}^{(+\nu)}\oplus\mathcal{H}^{(-\nu)}$ can be constructed from two two-band models $\mathcal{H}^{(\pm \nu)}$ which have the same gap. In this situation, the effective-band-gap can close for large enough magnetic field (the case for $\nu=1$ is discussed in Ref. \onlinecite{zhang16} and see Supplemental Material for the case of $\nu=2$).
To describe the gap movement, we note that for two (nearly) parabolic bands $i$ and $j$ with effective mass inverse $2\beta$ and $-2\beta'$, like in Eq. \eqref{HEone}, the crossing energy inside magnetic field is estimated as
\begin{equation}
\label{eq:crossingenergy}
    E_{c}=\frac{2\beta\beta'(\mathcal{Q}_{ii}-\mathcal{Q}_{jj})}{\beta+\beta'}B,
\end{equation}
neglecting the deformation of bands caused by the off-diagonal entries. Therefore if one band is flat, the crossing energy (and the gap) remains at zero and there is generally no in-gap ALLs \cite{SM}.

\begin{figure*}[t]
\centering
\includegraphics[width=\textwidth]{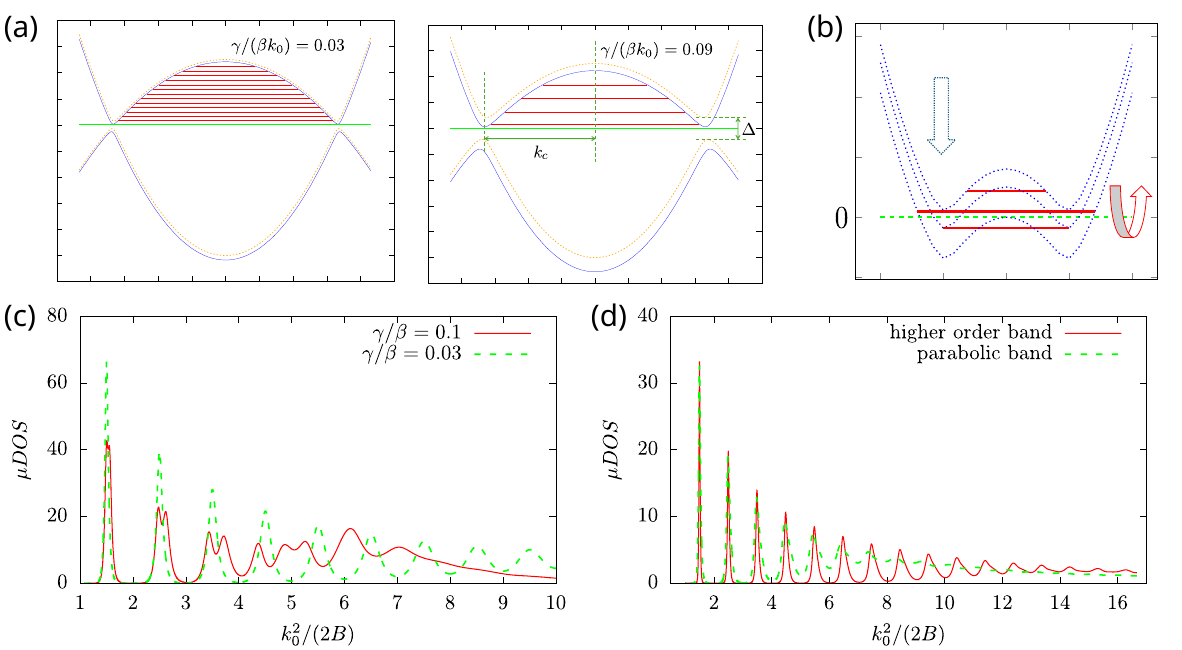}
\caption{\textbf{QO from effective band picture}. 
(a) The LLs that are going to cross the chemical potential at critical magnetic field $B_{c}$ for two-band model $\mathcal{H}_{E}^{(-1)}$ with $\frac{\beta'}{\beta}=0.5$. From left to right $\frac{\gamma}{\beta k_{0}}=0.03$ and $0.09$ respectively. The dotted lines represent the original band at zero field, the gap $\Delta$ and critical momentum $k_{c}$ is shown for $\frac{\gamma}{\beta k_{0}}=0.09$. The former has larger $\tilde{N}_{LL}$. (b) One single LL (red line) crosses the chemical potential (green dashed line) twice with the changing magnetic field, while the effective band is moving in one direction. (c) Quantum oscillation ($\mu$DOS in arbitrary unit versus the inverse of magnetic field) for $\mathcal{H}^{(+2)}\oplus\mathcal{H}^{(-2)}$ with $\frac{\beta'}{\beta}=0.5$, in-gap chemical potential is set at $\frac{\mu}{\beta k_{0}^{2}}=0$ and the temperature $\frac{k_{B}T}{\beta k_{0}^{2}}=0.01$. The red solid line is for bigger gap $\frac{\gamma}{\beta}=0.1$ and the green dashed line is for smaller gap $\frac{\gamma}{\beta}=0.03$. (d) Quantum oscillation ($\mu$DOS in arbitrary unit versus the inverse of magnetic field) for $\mathcal{H}^{(+1)}$ with $\frac{\beta'}{\beta}=0.5$, $\frac{\gamma}{\beta k_{0}}=0.04$, in-gap chemical potential is set at $\frac{\mu}{\beta k_{0}^{2}}=0$ and the temperature $\frac{k_{B}T}{\beta k_{0}^{2}}=0.01$. The red solid line is for higher-order (quartic) diagonal terms with $\frac{\alpha k_{0}^{2}}{\beta}=0.5$,  $\frac{\alpha' k_{0}^{2}}{\beta}=0.2$. The green dashed line is for parabolic diagonal terms with $\alpha=\alpha'=0$.}
\label{Fig:QO}
\end{figure*}

From the effective-band picture, it is clear that the QO from ALLs is determined by two factors with varying magnetic field: (i) the broadening of LL gaps inside the effective bands, this corresponds to what the normal QO measures in metals; (ii) the movement of effective bands themselves. Here we take $\mathcal{H}^{(-1)}$ (with diagonal piece being parabolic bands) as an example for illustrating the mechanism. As shown in Fig. \ref{Fig:EB} (a), the upper effective band is moving down towards the chemical potential at energy zero. The LLs that are moving down with the upper-effective-band are those with small quantum numbers, namely the ones in the ``lump" in the middle of the band. At the critical magnetic field $B_{c}$ the upper-effective-band reaches (tangent to) the chemical potential, as illustrated in Fig. \ref{Fig:QO} (a). After that, all the LLs left in the ``lump" are going to cross the chemical potential. As shown in Fig. \ref{Fig:QO} (b), each of these LLs will cross the chemical potential twice since they eventually go up. 

Assuming the gap is $\Delta$, with Eq. \eqref{eq:crossingenergy} we can have an estimate for $B_{c}$, namely $B_{c}=\frac{(\beta+\beta')\Delta}{4\beta\beta'(\mathcal{Q}_{ii}-\mathcal{Q}_{jj})}$. At $B_{c}$, if the band touches the chemical potential with critical momentum $k_{c}$, then the number of LLs that are going to cross the chemical potential is given by $\tilde{N}_{LL}\sim \frac{k_{c}^{2}}{2B_{c}}$. Using the estimation before we arrive at the following formula for the available LLs (excluding those from extra ansatz states),
\begin{equation}
\label{eq:NLL}
    \tilde{N}_{LL}=\frac{2\beta\beta'(\mathcal{Q}_{ii}-\mathcal{Q}_{jj})k_{c}^{2}}{(\beta+\beta')\Delta}.
\end{equation}
When $\tilde{N}_{LL}\gg 1$, there are plenty of LLs left in the ``lump", thus we expect to observe fermi-surface-like QO in this limit. On the other hand for $\tilde{N}_{LL}\sim 1$, the system is already in the quantum limit at $B_{c}$ and so the QO observed only has a few peaks and does not have a clear period. For the case of Hamiltonian $\mathcal{H}^{(\pm 1)}$ given in Eq. \eqref{HEone}, we estimate $k_{c}\approx k_{0}$, $\Delta\sim \gamma k_{0}$ and we take $\frac{\beta'}{\beta}=0.5$, so in this case $\tilde{N}_{LL}\approx \frac{2}{3}(\frac{\gamma}{\beta k_{0}})^{-1}$. In Fig. \ref{Fig:QO} (a), we plot the situation at $B_{c}$ for two values of $\gamma$ (for $\frac{\gamma}{\beta k_{0}}=0.03$, the estimation is $\tilde{N}_{LL}\approx 22$ and for $\frac{\gamma}{\beta k_{0}}=0.09$, the estimation is $\tilde{N}_{LL}\approx 7$), as shown, the formula provides a good estimation on the available LLs that are going to cross the chemical potential.

To characterize the QO from ALLs, we focus on the quantity $\mu-$density of states ($\mu DOS$) \cite{zhang16,cooper17,shen18}, defined as the response of total number of particle under the change of chemical potential in a grand canonical ensemble: 
\begin{equation}
    \mu DOS=\sum_{n,\alpha}\frac{\partial n_{F}(\epsilon_{n}^{\alpha},\mu)}{\partial\mu}=\mathcal{D}\sum_{n=0}^{\infty}\frac{1}{2k_{B}T}\frac{1}{1+\cosh(\frac{E_{n}-\mu}{k_{B}T})}.
\end{equation}
This quantity can reflect the conductivity of the SdH effect. In Fig. \ref{Fig:QO} (c) we plot the QO of $\mu DOS$ with respect to the inverse of magnetic field at low temperature for $\mathcal{H}^{(+2)}\oplus\mathcal{H}^{(-2)}$ with a smaller gap  ($\frac{\gamma}{\beta}=0.03$) and a bigger gap ($\frac{\gamma}{\beta}=0.1$); the former case is in the large $\tilde{N}_{LL}$ limit while the latter approaches the quantum limit (See Supplemental Material \cite{SM} for more details of the analysis and QO in the three-band model (\ref{threebm})).

Furthermore, the effect of higher-order terms in the diagonal of the Hamiltonian can be studied. 
Here we take one example for $\mathcal{H}^{(+1)}$ with diagonal piece being quartic in $k$ (the diagonal piece for (\ref{Hk1}) reads $\text{diag}[\alpha(k^{4}-k_{0}^{4})+\beta(k^{2}-k_{0}^{2}),-\alpha'(k^{4}-k_{0}^{4})-\beta'(k^{2}-k_{0}^{2})]$), and compare the QO for the case with finite $\alpha$ (quartic band) and vanishing $\alpha$ (parabolic band). 
A finite $\alpha$ lowers the critical field $B_c$ at which QO begins. This leads to a larger field range for QO before the quantum limit, rendering the QO more pronounced.
The result is shown in Fig. \ref{Fig:QO} (d), from which one can clearly see that including higher-order diagonal terms results in more peaks in the QO while keeping other parameters fixed. 

What does the effective-band picture tell about the frequency (period) of the QO? To answer this question, we first note that the ``fermi surface" of the effective-band can be obtained from the effective Hamiltonian $\mathcal{H}_{E}(k_{\xi},B)$ by eigenvalue equation $\varepsilon(k_{\xi},B)=0$, assuming chemical potential is at energy zero. Such equation leads to the relation $k_{\xi}(B)$ and further determines $\xi$ as a function of $B$ through Eq. \eqref{orbitalkn}. Whenever $\xi$ coincides with one from the sequence of the quantization rule, there is a peak for the $\mu DOS$. The frequency $f$ of the oscillation can be obtained as 
\begin{equation}
\label{QOfrequency}
    f\equiv\frac{d\xi}{d(\frac{1}{B})}=\frac{\mathcal{S}_{E}}{2\pi}+\frac{1}{2}B\frac{\frac{\partial\varepsilon}{\partial B}}{\frac{\partial \varepsilon}{\partial k^{2}}},
\end{equation}
in which $\mathcal{S}_{E}=\pi k_{\xi}^{2}=2\pi B(\xi+\frac{1}{2})$ is the area of the effective-band fermi surface \cite{SM}. This equation shows that the frequency of the ALL QO is not determined by the fermi surface area of the effective-band; rather, there is a modification. When $\varepsilon$ has no explicit dependence on $B$, we restore the Lifshitz-Onsager result. Such formula can explain the splitting of QO peaks when gap is larger and the system is reaching the quantum limit. Such effect is more pronounced for gap-sharing two-component models $\mathcal{H}^{(+\nu)}\oplus\mathcal{H}^{(-\nu)}$, as is evident from Fig. \ref{Fig:QO} (c). For larger gaps there are two Fermi surfaces corresponding to two frequencies by \eqref{QOfrequency}; on the contrary, when gap is small, these two frequencies determined by \eqref{QOfrequency} converges.

We end this section by considering how the effective-band picture and ALL induced QO might be applicable for understanding in-gap QO reported in experiments, in parallel with other existing scenarios.
For the Kondo insulator $\text{SmB}_{6}$, fitting with experiments \cite{zhang16} reveals $\frac{\beta'}{\beta}\approx 0.1$ and $\frac{(\beta+\beta')k_{c}^{2}}{\Delta}\approx 42$, which gives $\tilde{N}_{LL}\approx 7$ from Eq.\ \eqref{eq:NLL}, which suggest in-gap QO is expected from the effective-band picture.
In comparison, for $\text{InAs/GaSb}$ quantum well \cite{HanZ19}, the number of available LLs including self-energy correction of the two-dimensional electron gas \cite{shen18} is $\tilde{N}_{LL}\approx 1$ in the inverted narrow-gap regime. This estimate might explain why fewer in-gap QO peaks were experimentally seen in this system.

\section{Discussions}

In this work we discussed a framework for studying LLs in rotationally invariant low-energy models, which we show are classified by a set of integers labeling the relative angular momentum among the bands. Starting from a general tight-binding Hamiltonian, one first obtains its low-energy Hamiltonian; if it has rotation symmetry, then a set of effective bands and quantization rules can be constructed based on its indices, using which its exact LLs can be obtained. The moving effective bands with magnetic field illustrate the evolvement of LLs, some of which can protrude into the original gap and become anomalous LLs. Specifically for low-energy models in canonical form, there are two conditions for the upmost filled band and the lowest empty band, namely (i) the diagonal pieces have to be dispersive (no flat band), (ii) there has to be relative angular momentum between the two bands. Using the effective-band picture we discuss examples of ALLs and analyze possible QO from ALLs in insulating models. In particular, we propose a formula (Eq. \eqref{eq:NLL}) for the available ALLs that are going to cross the chemical potential for ALL induced QO. Together with the estimate of critical magnetic field $B_{c}$, these provide a useful toolkit to determine whether ALL scenario can explain the QO in insulators.

The moving effective bands are capable of explaining more phenomena in quantum oscillation. For QO in metallic systems, the relative angular momenta between bands lead to observable effects. In particular, QO measured in organic conductors $\alpha-(\text{BEDT-TTF})_{2}I_{3}$ \cite{Tisserond17} and the conducting interfaces of $\text{LaVO}_{3}-\text{KTaO}_{3}$, $\text{EuO}-\text{KTaO}_{3}$ etc show frequencies deviated from the $1/B$ law; such QO is thus dubbed ``aperiodic quantum oscillation" \cite{Kumari25,Rubi20,Rubi24}. These metallic systems can be modeled by Hamiltonians with relative angular momenta between bands and chemical potential lying inside one band. The effective-band picture can offer clear description of such phenomena. With the changing effective Fermi surfaces, the QO frequencies are given by Eq. \eqref{QOfrequency} rather than the $1/B$ law.

For models with ALLs, one natural question to ask is whether this picture can be applied to low-energy models without rotational invariance. Here, we discuss a simple two-band model without continuous rotational symmetry which can still be treated within our framework.
The Hamiltonian of the two-band model is given by
\begin{equation}
\label{HkW}
\tilde{\mathcal{H}}_{\boldsymbol{k}}^{(\pm 1)}=\left(\begin{array}{cc}
\beta(k^{2}-k_{0}^{2})&\gamma k_{\mp}+W\\
\gamma k_{\pm}+W&-\beta'(k^{2}-k_{0}^{2})
\end{array}\right).
\end{equation}
For the case $\beta'=0$, its LL spectrum can be analytically obtained by both the number-basis calculation and the effective band picture with quantization rules (see Supplemental Materials \cite{SM} for details). For the effective-band picture we first use a generalized Galilean transformation to go to another reference frame going in the $x$-direction, the new momentum and energy satisfy $k_{x}=k_{x}'+\delta$, $k_{y}=k_{y}'$ and $E=E'$ in which $\delta=\delta(E)$ can be a function of energy. In the new reference frame, the Hamiltonian $\tilde{\mathcal{H}}'(\boldsymbol{k}')$ has rotational invariance when $\delta=-\frac{W\gamma}{\gamma^{2}+\beta E}$. At this specific choice of $\delta$, the language of effective bands can be applied; the quantization rules can be written in the following form $\left (k_{+}+\frac{W\gamma}{\gamma^{2}+\beta E} \right) \left(k_{-}+\frac{W\gamma}{\gamma^{2}+\beta E}\right)=2nB$, with $n=1,2,\cdots$. Using these one reproduces the exact LLs \cite{SM}. To further understand how this generalizes to models without solvable LLs, we study numerically the model (\ref{HkW}) with finite $\beta'$. Although quantization rules cannot be specified in this case, we find that effective band can still play a role revealing the ALLs (see Supplemental Materials \cite{SM} for details). 
Going beyond, a general conclusion on how the effective-band picture could be applicable to models without an exactly solvable limit is open, and it might be fruitful to consider models for which the dependence of the LL spectrum on $B$ is known although the models are not exactly solvable\cite{dietl08,kishigi17}.

For the discussion on QO in this work, we have inferred the oscillation from the $\mu DOS$, which can be attributed to the crossing of LLs with the chemical potential. Implicitly, this assumes the system can be described in the grand canonical potential. 
One might ask how QO could appear in alternative treatments where the particle number is held fixed. For metals, it was known theoretically that the magnetization of the system exhibits QO when particle number is fixed \cite{Shoenberg1984,Xu2008}. 
For a band insulator, however, one might argue that since the particle number is fixed by that of the fully filled bands, it is unclear if ALLs (or more generally, all LLs) could lead to QO. To this end, numerical studies on Hofstadter model have produced positive results \cite{Kishigi14}. Besides, another partial justification to the in-gap QO could be argued through the inevitable existence of in-gap disorder states, which could stabilize the chemical potential and therefore support QO in $\mu DOS$ as a proxy for the QO of physical observables. Alternatively, in systems with an indirect gap, in-gap QO could also result from the compensating behavior of states originating from different pockets in the Brillouin zone. The extent to which our present approach could be related to the experimental observed QOs in various insulating systems is an interesting open question.

\section*{Acknowledgements}
We thank Niels Schr\"oter and Yoonseok Hwang for helpful discussions.
This work is supported by by the National Key R \& D Program of China (Grant No. 2021YFA1401500) and the Hong Kong Research Grant Council (ECS 26308021 and C7037-22GF).



\bibliography{refALL}

\clearpage
\widetext
\begin{center}
{\Large \textbf{Supplemental Materials}}
\end{center}
\setcounter{equation}{0}
\setcounter{figure}{0}
\setcounter{table}{0}
\setcounter{page}{1}
\makeatletter
\renewcommand{\theequation}{S\arabic{equation}}
\renewcommand{\thefigure}{S\arabic{figure}}

\section*{Analysis on the three-band models}

For three-band models, we need the group $SU(3)$ and the eight Gell-Mann matrices 
\begin{eqnarray}
\begin{aligned}
&\lambda_{1}=\left(\begin{array}{ccc}
0&1&0\\
1&0&0\\
0&0&0
\end{array}\right),\quad \lambda_{2}=\left(\begin{array}{ccc}
0&-i&0\\
i&0&0\\
0&0&0
\end{array}\right),\quad
\lambda_{3}=\left(\begin{array}{ccc}
1&0&0\\
0&-1&0\\
0&0&0
\end{array}\right),
\lambda_{4}=\left(\begin{array}{ccc}
0&0&1\\
0&0&0\\
1&0&0
\end{array}\right),\\ 
&\lambda_{5}=\left(\begin{array}{ccc}
0&0&-i\\
0&0&0\\
i&0&0
\end{array}\right),\quad
\lambda_{6}=\left(\begin{array}{ccc}
0&0&0\\
0&0&1\\
0&1&0
\end{array}\right),\quad
\lambda_{7}=\left(\begin{array}{ccc}
0&0&0\\
0&0&-i\\
0&i&0
\end{array}\right),\quad \lambda_{8}=\frac{1}{\sqrt{3}}\left(\begin{array}{ccc}
1&0&0\\
0&1&0\\
0&0&-2
\end{array}\right).
\end{aligned}
\end{eqnarray}
The generators of $SU(3)$ are $\hat{T}_{a}=\frac{1}{2}\lambda_{a}$. Following usual nomenclature we define the following operators $I_{\pm}=\hat{T}_{1}\pm i\hat{T}_{2}$, $U_{\pm}=\hat{T}_{6}\pm i\hat{T}_{7}$, $V_{\pm}=\hat{T}_{4}\pm i\hat{T}_{5}$, $I_{3}=\hat{T}_{3}$ and $Y=\frac{2}{\sqrt{3}}\hat{T}_{8}$. The Lie algebra of $SU(3)$ is given by
\begin{eqnarray}
\label{SU3Lie}
\begin{aligned}
&[I_{3},I_{\pm}]=\pm I_{\pm},\quad [I_{3},U_{\pm}]=\mp \frac{1}{2}U_{\pm},\quad [I_{3},V_{\pm}]=\pm \frac{1}{2}V_{\pm},\quad [Y,I_{\pm}]=0,\\
&[Y,U_{\pm}]=\pm U_{\pm},\quad [Y,V_{\pm}]=\pm V_{\pm}, \quad [I_{+},I_{-}]=2 I_{3},\quad [I_{+},V_{-}]=-U_{-},\quad [I_{+},U_{+}]=V_{+},\\
&[U_{+},V_{-}]=I_{-}, \quad [I_{+},V_{+}]=[I_{+},U_{-}]=[U_{+},V_{+}]=0.\\
&[U_{+},U_{-}]=\frac{3}{2}Y-I_{3}=\sqrt{3}\hat{T}_{8}-\hat{T}_{3}=2U_{3},\quad [V_{+},V_{-}]=\frac{3}{2}Y+I_{3}=\sqrt{3}\hat{T}_{8}+\hat{T}_{3}=2V_{3}.
\end{aligned}
\end{eqnarray}
Among these, $\hat{I}_{3}=\frac{1}{2}\text{diag}(1,-1,0)$ and $\hat{Y}=\frac{1}{3}\text{diag}(1,1,-2)$ generate the $U(1)$ subgroups. The generator $\hat{\mathcal{Q}}=p\hat{I}_{3}+(q+\frac{p}{2})\hat{Y}=\text{diag}(\frac{2p}{3}+\frac{q}{3}, -\frac{p}{3}+\frac{q}{3},-\frac{p}{3}-\frac{2q}{3})$ with two integers $(p,q)$ characterizing the relative angular momenta. The general three-band Hamiltonian reads
\begin{equation}
\label{generalthreeband}
\mathcal{H}_{\boldsymbol{k}}=\hat{D}(k)+\left(u(\boldsymbol{k})\hat{U}_{+}+v(\boldsymbol{k})\hat{V}_{+}+i(\boldsymbol{k})\hat{I}_{+}+\text{h.c.}\right),
\end{equation}
in which $\hat{D}(k)=h_{0}\hat{I}+h_{1}\hat{Y}+h_{3}\hat{I}_{3}$ is the diagonal piece. For infinitesimal rotation, the adjoint action \eqref{adjointaction}
and the Lie algebra of $SU(3)$ \eqref{SU3Lie} imply the following transformation of the components: $i(\boldsymbol{k})\rightarrow (1+i\theta p)i(\boldsymbol{k})$, $u(\boldsymbol{k})\rightarrow (1+i\theta q)u(\boldsymbol{k})$ and $v(\boldsymbol{k})\rightarrow (1+i\theta(p+q))v(\boldsymbol{k})$. Therefore the canonical Hamiltonian is given by
\begin{equation}
\mathcal{H}_{\boldsymbol{k}}^{(p,q)}=\left(\begin{array}{ccc}
\cdots&\gamma_{1}(k_{-})^{p}&\gamma_{2}(k_{-})^{p+q}\\
\gamma_{1}(k_{+})^{p}&\cdots&\gamma_{3}(k_{-})^{q}\\
\gamma_{2}(k_{+})^{p+q}&\gamma_{3}(k_{+})^{q}&\cdots
\end{array}\right).
\end{equation}

\section*{Details of the effective-band picture and extra ansatz states}

\subsection*{Two-band model $\mathcal{H}^{(\pm 2)}$}

For two-band models $\mathcal{H}_{\boldsymbol{k}}^{(\pm 2)}$, the U(1) generator $\hat{\mathcal{Q}}=\pm\sigma^{z}$, from which we obtain the quantization rule: $\xi=1, 1+1,2+1,\cdots$, here we take the diagonal pieces to be parabolic bands. The only off-diagonal term has $r=2$, $s=0$, so the effective Hamiltonian is 
\begin{equation}
\label{Hsb2}
\mathcal{H}_{E}^{(\pm 2)}=\left(\begin{array}{cc}
\beta(k^{2}-k_{0}^{2})\pm 2B\beta&\gamma\sqrt{k^{4}-B^{2}}\\
\gamma\sqrt{k^{4}-B^{2}}&-\beta'(k^{2}-k_{0}^{2})\pm 2B\beta'
\end{array}\right).
\end{equation}  
For $\mathcal{H}^{(2)}$ the magnetic Hamiltonian is
\begin{equation}
\mathcal{H}^{(2)}_{B}=(2B)\left(\begin{array}{cc}
\beta[a^{\dagger}a+\frac{1}{2}-\frac{k_{0}^{2}}{2B}]&\gamma (a^{\dagger})^{2}\\
\gamma a^{2}&-\beta'[a^{\dagger}a+\frac{1}{2}-\frac{k_{0}^{2}}{2B}]
\end{array}\right).
\end{equation}
So it has two extra ansatz states: $(|0\rangle, 0)^{T}$ with energy $\beta(B-k_{0}^{2})$ and $(|1\rangle,0)^{T}$ with energy $\beta(3B-k_{0}^{2})$.

From the two-band model $\mathcal{H}^{(\pm 2)}$ one can construct four-band model $\mathcal{H}^{(+2)}\oplus\mathcal{H}^{(-2)}$, whose effective band is given by
\begin{equation}
    \mathcal{H}_{E}^{(+2)}\oplus\mathcal{H}_{E}^{(-2)}=\left(\begin{array}{cccc}
    \beta(k^{2}-k_{0}^{2})+ 2B\beta&\gamma\sqrt{k^{4}-B^{2}}&0&0\\
     \gamma\sqrt{k^{4}-B^{2}}&-\beta'(k^{2}-k_{0}^{2})- 2B\beta'&0&0\\
     0&0&\beta(k^{2}-k_{0}^{2})- 2B\beta&\gamma\sqrt{k^{4}-B^{2}}\\
     0&0&\gamma\sqrt{k^{4}-B^{2}}&-\beta'(k^{2}-k_{0}^{2})+ 2B\beta'
    \end{array}\right)
\end{equation}

\begin{figure}
\includegraphics[width=0.6\textwidth]{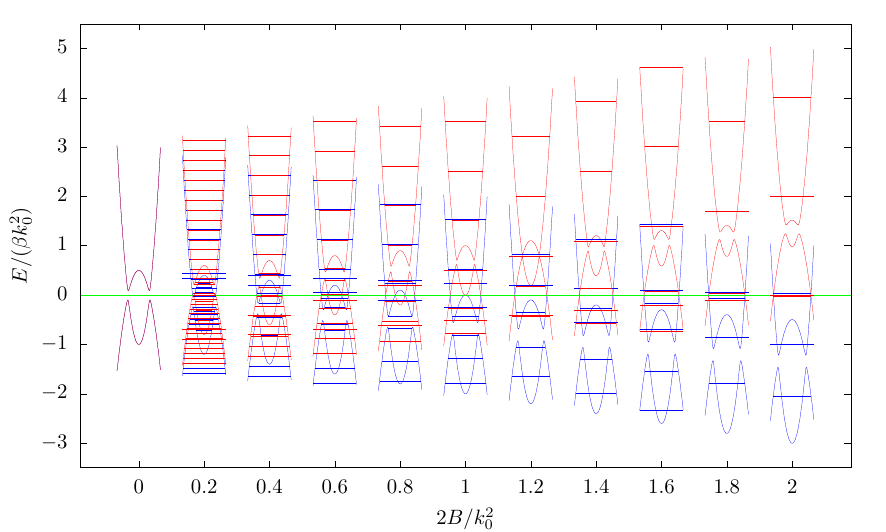}
\caption{Effective band evolution under magnetic field for four-band $\mathcal{H}^{(+2)}\oplus\mathcal{H}^{(-2)}$ with $\frac{\beta'}{\beta}=0.5$ and $\frac{\gamma}{\beta}=0.1$. The left-most band is the original band for $B=0$. The chemical potential is set at $\frac{\mu}{\beta k_{0}^{2}}=0$.}
\label{figsbs1}
\end{figure}

Its effective-band is shown in Fig. \ref{figsbs1}, from which we can see its effective band-gap closes as magnetic field $B$ gets large.

\subsection*{Three-band model $\mathcal{H}_{\boldsymbol{k}}^{(1,-1)}$}

For three-band model $\mathcal{H}_{\boldsymbol{k}}^{(1,-1)}$ (\ref{threebm}), the magnetic Hamiltonian is given by
\begin{equation}
\mathcal{H}^{(1,-1)}=\left(\begin{array}{ccc}
\beta_{1}[2B(a^{\dagger}a+\frac{1}{2})+k_{1}^{2}]&\gamma_{1}\sqrt{2B}a^{\dagger}&\gamma_{2}k_{0}\\
\gamma_{1}\sqrt{2B}a&\beta[2B(a^{\dagger}a+\frac{1}{2})-k_{0}^{2}]&\gamma_{3}\sqrt{2B}a\\
\gamma_{2}k_{0}&\gamma_{3}\sqrt{2B}a^{\dagger}&-\beta_{2}[2B(a^{\dagger}a+\frac{1}{2})-k_{0}^{2}]
\end{array}\right).
\end{equation}
The ansatz solution is given by $(|n\rangle, |n-1\rangle, |n\rangle)^{T}$, with $n=1,2,\cdots$. 

On the other hand, the effective band is given by Eq. \eqref{threebsb}, it is governed by the following dimensionless quantities: $\frac{\beta_{1}}{\beta}$, $\frac{\beta_{2}}{\beta}$, $\frac{k}{k_{0}}$, $\frac{k_{1}}{k_{0}}$, $\frac{2B}{k_{0}^{2}}$, $\frac{\gamma_{1}}{\beta k_{0}}$, $\frac{\gamma_{2}}{\beta k_{0}}$, $\frac{\gamma_{3}}{\beta k_{0}}$. In terms of these the effective Hamiltonian can be written as
\begin{equation}
\mathcal{H}_{E}^{(1,-1)}=(\beta k_{0}^{2})\left(\begin{array}{ccc}
\frac{\beta_{1}}{\beta}[(\frac{k}{k_{0}})^{2}+(\frac{k_{1}}{k_{0}})^{2}]+\frac{\beta_{1}}{\beta}(\frac{2B}{k_{0}^{2}})(\frac{1}{3})&\frac{\gamma_{1}}{\beta k_{0}}\sqrt{(\frac{k}{k_{0}})^{2}-\frac{2B}{k_{0}^{2}}\frac{1}{6}}&\frac{\gamma_{2}}{\beta k_{0}}\\
\frac{\gamma_{1}}{\beta k_{0}}\sqrt{(\frac{k}{k_{0}})^{2}-\frac{2B}{k_{0}^{2}}\frac{1}{6}}&(\frac{k}{k_{0}})^{2}-1+\frac{2B}{k_{0}^{2}}(-\frac{2}{3})&\frac{\gamma_{3}}{\beta k_{0}}\sqrt{(\frac{k}{k_{0}})^{2}-\frac{2B}{k_{0}^{2}}\frac{1}{6}}\\
\frac{\gamma_{2}}{\beta k_{0}}&\frac{\gamma_{3}}{\beta k_{0}}\sqrt{(\frac{k}{k_{0}})^{2}-\frac{2B}{k_{0}^{2}}\frac{1}{6}}&-(\frac{\beta_{2}}{\beta})[(\frac{k}{k_{0}})^{2}-1]-(\frac{\beta_{2}}{\beta})(\frac{2B}{k_{0}^{2}})\frac{1}{3}
\end{array}\right).
\end{equation}
It has one extra ansatz solution $(|0\rangle, 0,|0\rangle)^{T}$, correspondingly there are two LLs whose energies are eigenvalues of the two-by-two matrix
\begin{equation}
\mathcal{H}_{\text{ex}}=\left(\begin{array}{cc}
\beta_{1}(B+k_{1}^{2})&\gamma_{2}k_{0}\\
\gamma_{2}k_{0}&-\beta_{2}(B-k_{0}^{2})
\end{array}\right).
\end{equation}
The energies of extra LLs are given by
\begin{equation}
E_{\text{ex}}=\frac{1}{2}\left[\beta_{1}(B+k_{1}^{2})-\beta_{2}(B-k_{0}^{2})\pm\sqrt{[\beta_{1}(B+k_{1}^{2})+\beta_{2}(B-k_{0}^{2})]^{2}+4\gamma_{2}^{2}k_{0}^{2}}\right].
\end{equation}

\begin{figure}
\includegraphics[width=0.6\textwidth]{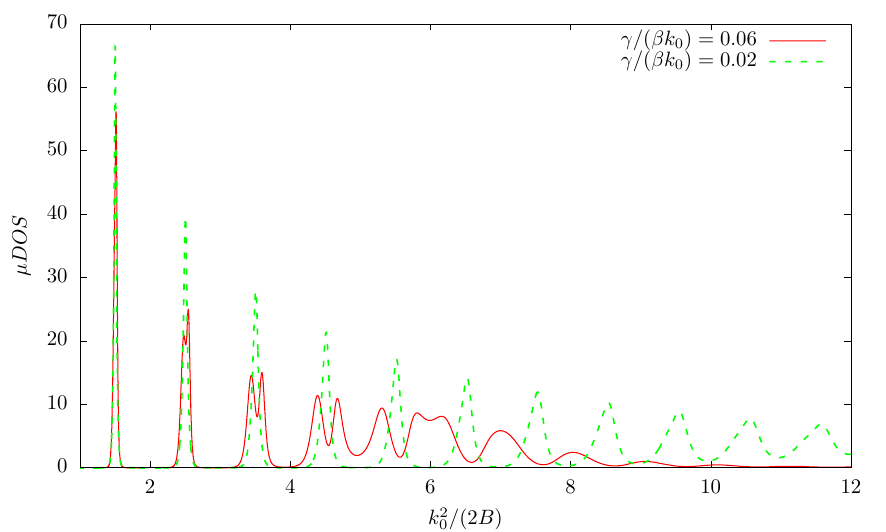}
\caption{Quantum oscillation ($\mu$DOS in arbitrary unit versus the inverse of magnetic field) for $\mathcal{H}^{(1,-1)}$ with $\frac{\beta_{1}}{\beta}=1.2$, $\frac{\beta_{2}}{\beta}=0.5$, $\frac{k_{1}}{k_{0}}=0.5$, in-gap chemical potential is set at $\frac{\mu}{\beta k_{0}^{2}}=0$ and the temperature $\frac{k_{B}T}{\beta k_{0}^{2}}=0.005$. The red solid line is for bigger gap with $\frac{\gamma_{1}}{\beta k_{0}}=\frac{\gamma_{2}}{\beta k_{0}}=\frac{\gamma_{3}}{\beta k_{0}}=0.06$, the green dashed is for smaller gap with $\frac{\gamma_{1}}{\beta k_{0}}=\frac{\gamma_{2}}{\beta k_{0}}=\frac{\gamma_{3}}{\beta k_{0}}=0.02$.}
\label{figQO2}
\end{figure}

In Fig. \ref{figQO2} we plot the QO of $\mu$DOS for the three-band model $\mathcal{H}^{(1,-1)}$ with parameters used in Fig. \ref{Fig:EB} for the bigger-gap case and a small-gap case with Fermi-surface-like behavior.

\subsection*{Three-band model $\mathcal{H}_{\boldsymbol{k}}^{(-1,-1)}$}

Following the discussion of the Lieb lattice model with SOC, we consider the effective model given by Eq. (\ref{Hthreeone}) in detail. The magnetic Hamiltonian is given by
\begin{equation}
    \mathcal{H}_{B}^{(-1,-1)}=\left(\begin{array}{ccc}
    \lambda[4-B(a^{\dagger}a+\frac{1}{2})]&i\sqrt{B}a &0  \\
     -i\sqrt{B}a^{\dagger}&0&i\sqrt{B}a\\
     0&-i\sqrt{B}a^{\dagger}&\lambda[B(a^{\dagger}a+\frac{1}{2})-4]
    \end{array}
    \right).
\end{equation}
It has two extra ansatz states, $(0,|0\rangle,|1\rangle)^{T}$ and $(0,0,|0\rangle)^{T}$. The former corresponds to two extra LLs, whose energies are eigenvalues of
\begin{equation}
    \left(\begin{array}{cc}
        0 & i\sqrt{B} \\
        -i\sqrt{B} & \lambda(\frac{3}{2}B-4) 
    \end{array}
    \right),
\end{equation}
and are given by $E_{1,2}=\frac{1}{2}[\lambda(\frac{3}{2}B-4)\pm\sqrt{\lambda^{2}(\frac{3}{2}B-4)^{2}+4B}]$. The latter has one extra LL with the energy $E_{3}=\lambda(\frac{1}{2}B-4)$.

\subsection*{ALLs in flat band touching with parabolic band}

\begin{figure}
\includegraphics[width=0.5\textwidth]{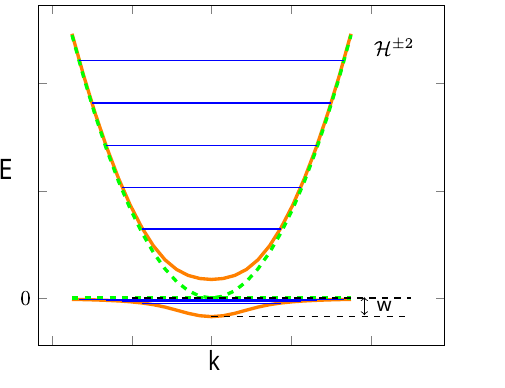}
\caption{Effective band of the kagome lattice model $\mathcal{H}^{(\pm 2)}$, green dashed represents the original band, blue lines are the ALLs of the lower band. The band width of the lower band is labeled as $w$.}
\label{figkagome}
\end{figure}

The effective-band picture can be used to study ALLs of flat bands. Besides the isolated-flat-band case discussed in the main text,
another situation is flat band that touches with another dispersive band. In such systems, flat bands are categorized into singular and non-singular flat bands based on the continuity of wavefunction around the band-touching point \cite{Yang20}. With the indices of models, we show that the distinction of singular flat bands from non-singular flat bands is that the indices of the flat band and the parabolic band are different; in other words, physically a singular flat band has non-zero angular momentum difference from the parabolic band it touches to. To see this we revisit the electron hopping model on Kagome lattice \cite{guo09}. Among the three bands, one has high energy and is isolated from the other two. After projected onto the two low-energy bands, the Hamiltonian of the low-energy effective model is given by \cite{Yang20}
\begin{equation}
\mathcal{H}_{\text{kagome}}=\left(\begin{array}{cc}
\frac{k_{x}^{2}}{4}&-\frac{ik_{x}k_{y}}{4}\\
\frac{ik_{x}k_{y}}{4}&\frac{k_{y}^{2}}{4}
\end{array}\right)
\end{equation} 
Using a constant unitary transformation $U=\frac{1}{\sqrt{2}}(\hat{\sigma}_{z}+\hat{\sigma}_{x})$, with $\hat{\sigma}_{z,x}$ being Pauli matrices, the Hamiltonian can be brought into canonical form $\mathcal{H}^{(-2)}$, namely $U\mathcal{H}_{\text{kagome}}U^{\dagger}\rightarrow \mathcal{H}^{(-2)}$, this transformed Hamiltonian belongs to a general set of Hamiltonians $\mathcal{H}^{(\pm 2)}$ with band-touching,
\begin{equation}
\mathcal{H}^{(\pm 2)}=\left(\begin{array}{cc}
\beta k^{2}&\gamma k_{\mp}^{2}\\
\gamma k_{\pm}^{2}&\beta' k^{2}
\end{array}\right), 
\end{equation}
the condition for flat band is that $\beta\beta'=\gamma^{2}$. The effective band for this system is given by
\begin{equation}
\mathcal{H}_{E}^{(\pm 2)}=\left(\begin{array}{cc}
\beta(k^{2}\pm 2B)&\gamma\sqrt{k^{4}-B^{2}}\\
\gamma\sqrt{k^{4}-B^{2}}&\beta'(k^{2}\mp 2B).
\end{array}\right).
\end{equation}
The flat band becomes dispersive for nonzero $B$, the LLs are thus carried away and form a band with finite bandwidth, as shown in Fig. \ref{figkagome}. For the simplest case $\beta=\beta'=1$, the lower band with finite range is given by $E_{-}=\frac{-3B^{2}}{k^{2}+\sqrt{k^{4}+3B^{2}}}$. The width of the band can reveal the evolvement of (and thus provides a rough estimate of) the ``spread of LLs", it is given by $w=\sqrt{3}B$ for this case. For general parameters it is $w=B[\sqrt{\beta^{2}+\beta'^{2}+\beta\beta'}-(\beta-\beta')]$. For a wide range of models, it is shown that this quantity is related to the quantum distance of the eigenfunctions \cite{Yang20}. In contrast, for non-singular flat bands in $\mathcal{H}^{(0)}$, the bandwidth $w$ and the spread of LLs are identically zero.

\section*{Analysis of QO using the effective-band picture}

From the effective-band picture we can obtain some information about the ALLs. First let us assume that in a multi-band model two parabolic band $i$ and $j$ cross at $E=0$ with some hybridization among themselves and with other bands. The two parabolic bands are given by $\beta(k^{2}-k_{0}^{2})$ and $-\beta'(k^{2}-k_{0}^{2})$, with corresponding eigenvalue of $\hat{\mathcal{Q}}$ being $\mathcal{Q}_{ii}$ and $\mathcal{Q}_{jj}$. Inside magnetic field, approximately the new crossing point for these two bands is determined by 
\begin{equation}
\beta(k^{2}-k_{0}^{2})+\mathcal{Q}_{ii}(2\beta)B=-\beta'(k^{2}-k_{0}^{2})-\mathcal{Q}_{jj}(2\beta')B.
\end{equation}
The new crossing energy for the two bands is given by
\begin{equation}
E_{c}=\frac{2\beta\beta'(\mathcal{Q}_{ii}-\mathcal{Q}_{jj})}{\beta+\beta'}B.
\end{equation}
If one of these bands is flat, then the energy of the band crossing remains zero. Therefore flat-band case generally does not have {\it in-gap} ALLs. But for two-band models with one flat band, ALLs can appear in the half-infinite region with no states in the original spectrum.

On the other hand, taking the changing effective bands as ``reference frame", the LLs themselves move, leading to the broading of the LL gap. To quantify this we note that the {\it density of states} is defined by 
\begin{equation}
\mathcal{D}=\frac{2\pi k dk}{dE}.
\end{equation}
For the quantization rules defined by $k_{n}=\sqrt{2B(|\zeta|+n+\frac{1}{2})}$, $n=0,1,2,\cdots$ we have the density of states near the $n$th LL,
\begin{equation}
\mathcal{D}=2\pi(|\zeta|+n+\frac{1}{2})\frac{dB}{dE}.
\end{equation}
Therefore we have, in the effective-band reference frame, the change of the $n$th LL with respect to magnetic field
\begin{equation}
\frac{d\tilde{E}_{n}}{dB}=\frac{2\pi(|\zeta|+n+\frac{1}{2})}{\mathcal{D}}.
\end{equation}

As a rough estimate, neglecting the changing shape of the effective bands, we have the change of $n$th LL with respective of the magnetic field is given by
\begin{equation}
\label{estimateofLL}
\frac{dE_{n}}{dB}=\frac{d\tilde{E}_{n}}{dB}+\frac{dE_{c}}{dB}=\frac{2\pi(|\zeta|+n+\frac{1}{2})}{\mathcal{D}}+\frac{2\beta\beta'(\mathcal{Q}_{ii}-\mathcal{Q}_{jj})}{\beta+\beta'}.
\end{equation}

We then turn to the implication of the effective-band picture on the frequency of oscillation. We set the chemical potential to be at energy zero. The zero-energy solution of the effective Hamiltonian $\mathcal{H}_{E}(k_{\xi},B)$ gives a equation on $k_{\xi}$ and $B$, namely
\begin{equation}
    \varepsilon(k_{\xi},B)=0,
\end{equation}
the solution to this equation $k_{\xi}(B)$ determines the crossing $\xi$. Whenever this $\xi$ coincides with one from the quantization rule, there is a peak in the QO. As $k_{\xi}^{2}=2B(\xi+\frac{1}{2})$, we have 
\begin{equation}
    \frac{d\varepsilon}{d B}=\frac{\partial\varepsilon}{\partial k^{2}}\frac{\partial k^{2}}{\partial B}+\frac{\partial \varepsilon}{\partial B}=2\frac{\partial \varepsilon}{\partial k^{2}}(\xi+\frac{1}{2})+\frac{\partial\varepsilon}{\partial B},
\end{equation}
and 
\begin{equation}
    \frac{d \varepsilon}{d\xi}=\frac{\partial\varepsilon}{\partial k^{2}}\frac{\partial k^{2}}{\partial\xi}=2B\frac{\partial\varepsilon}{\partial k^{2}}.
\end{equation}
The frequency of the oscillation is determined by
\begin{equation}
    f=\frac{d\xi}{d(\frac{1}{B})}=B^{2}\frac{\frac{d\varepsilon}{dB}}{\frac{d\varepsilon}{d\xi}}=B(\xi+\frac{1}{2})+\frac{1}{2}B\frac{\frac{\partial\varepsilon}{\partial B}}{\frac{\partial\varepsilon}{\partial k^{2}}}.
\end{equation}
The cross section of the effective band has area $\mathcal{S}_{E}=\pi k^{2}_{\xi}$, so the frequency is also written as
\begin{equation}
   f=\frac{\mathcal{S}_{E}}{2\pi}+\frac{1}{2}B\frac{\frac{\partial\varepsilon}{\partial B}}{\frac{\partial \varepsilon}{\partial k^{2}}}.
\end{equation}

\section*{The model without rotational invariance}

Here, we discuss a simple two-band model without continuous rotational symmetry which can still be treated using the effective-band picture. The Hamiltonian of the two-band model is given by
\begin{equation}
\tilde{\mathcal{H}}_{\boldsymbol{k}}^{(\pm 1)}=\left(\begin{array}{cc}
\beta(k^{2}-k_{0}^{2})&\gamma k_{\mp}+W\\
\gamma k_{\pm}+W&-\beta'(k^{2}-k_{0}^{2})
\end{array}\right).
\end{equation}
The LLs for $\beta'=0$ can be obtained by the effective bands and quantization rules. Firstly we use a generalized Galilean transformation to go to another reference frame going in the $x$-direction, the new momentum and energy satisfy
\begin{equation}
\label{Galilean}
k_{x}=k_{x}'+\delta,\qquad k_{y}=k_{y}', \qquad E=E',
\end{equation}
in which $\delta=\delta(E)$ can be a function of energy. In the new reference frame, the Hamiltonian $\tilde{\mathcal{H}}'(\boldsymbol{k}')$ has rotational invariance when $\delta=-\frac{W\gamma}{\gamma^{2}+\beta E}$. At this specific choice of $\delta$, the language of effective bands can be applied; we first add the diagonal correction terms to the transformed Hamiltonian $\tilde{\mathcal{H}}'(\boldsymbol{k}')$, and obtain the quantization rules: $k'^{2}=2Bn$ with $n=1,2,\cdots$. Then using (\ref{Galilean}) with the specific function $\delta(E)$ one can obtain the effective Hamiltonian in the original momentum $k$, namely
\begin{equation}
\label{effectiveH}
\tilde{\mathcal{H}}_{E}^{(\pm 1)}=\left(\begin{array}{cc}
\beta (k^{2}-k_{0}^{2})\pm B\beta &\gamma k_{\mp}+W\\
\gamma k_{\pm}+W&0
\end{array}\right).
\end{equation}
And the quantization rules can be written in the following form
\begin{equation}
\label{quantizationrules}
\left (k_{+}+\frac{W\gamma}{\gamma^{2}+\beta E} \right) \left(k_{-}+\frac{W\gamma}{\gamma^{2}+\beta E}\right)=2nB, 
\end{equation}
with $n=1,2,\cdots$. Using these one reproduces the exact LLs \cite{SM}. The combination (\ref{effectiveH}) and (\ref{quantizationrules}) forms two functions for three variables $(k_{x},k_{y},E)$, the reason it can determine a series of discrete LLs is that a (generalized) angular variable becomes redundant, and only (generalized) radial variable has physical meaning. For small $W$ the model (\ref{HkW}) can be understood as a two-band model perturbed away from the rotationally invariant point. 

\subsection*{Number basis calculation for the case $\beta'=0$}

Here we look at the model without rotational invariance (\ref{HkW}) for the case $\beta'=0$,
\begin{equation}
\tilde{\mathcal{H}}_{k}^{(\pm 1)}=\left(\begin{array}{cc}
\beta(k^{2}-k_{0}^{2})&\gamma k_{\mp}+W\\
\gamma k_{\pm}+W&0
\end{array}\right),
\end{equation}
and solve its LL spectrum using standard method by translating $k_{\pm}$ into boson creation and annihilation operators.

The solution of its LL spectrum generally has the form $(|\psi^{\pm 1}\rangle,|\phi^{\pm 1}\rangle)^{T}$, one can eliminate $|\phi\rangle$ and get 
\begin{equation}
\label{H1psi}
\left[2B(\gamma^{2}+E\beta)(a^{\dagger}+\eta)(a+\eta)+\mathcal{P}^{\pm 1}(E,B)\right]|\psi^{\pm 1}\rangle=0,
\end{equation}
in which $E$ is the energy and $\eta=\frac{\gamma W}{\sqrt{2B}(\gamma^{2}+E\beta)}$, in the equation above,
\begin{equation}
\mathcal{P}^{+1}=W^{2}+E\beta(B-k_{0}^{2})-E^{2}-\frac{W^{2}\gamma^{2}}{\gamma^{2}+E\beta}
\end{equation}
and 
\begin{equation}
\mathcal{P}^{-1}=\mathcal{P}^{+1}+2B\gamma^{2}.
\end{equation}
Using (\ref{H1psi}) we find that the solution of $|\psi\rangle$ for both $\tilde{\mathcal{H}}_{n}^{(\pm 1)}$ is $|\psi^{\pm 1}\rangle=\hat{\Pi}|n\rangle=|\tilde{n}\rangle$, for $n=0,1,2,\cdots$, and the operator $\hat{\Pi}=e^{\eta a}e^{-\eta a^{\dagger}}$ is the translational operation on bosonic operators.

For $\tilde{\mathcal{H}}_{n}^{(+1)}$ in particular, the corresponding eigenvalues $E$ satisfy the following cubic equation
\begin{eqnarray}
\label{cubicplus}
\begin{aligned}
\beta E^{3}+\left(\gamma^{2}+\beta^{2}(k_{0}^{2}-B)-2B\beta^{2}n\right)E^{2}+\left(\gamma^{2}\beta(k_{0}^{2}-B)-W^{2}\beta-4B\beta \gamma^{2}n\right)E-2B\gamma^{4}n=0.
\end{aligned}
\end{eqnarray}
The $|\phi\rangle$ of $\tilde{\mathcal{H}}_{n}^{(+1)}$ can be written as $|\phi^{+1}\rangle=\frac{\gamma\sqrt{2Bn}}{E}(\hat{\Pi}|n-1\rangle)+\frac{W\beta}{E\beta+\gamma^{2}}(\hat{\Pi}|n\rangle)$. For $\tilde{\mathcal{H}}_{n}^{(-1)}$ the energies satisfy the following cubic equation
\begin{equation}
\label{Hminusonecubic}
\beta E^{3}+\left(\gamma^{2}+\beta^{2}(k_{0}^{2}-B)-\beta^{2}2Bn\right)E^{2}+\left(\gamma^{2}\beta(k_{0}^{2}-B)-W^{2}\beta-4B\gamma^{2}\beta(n+\frac{1}{2})\right)E-2B\gamma^{4}(1+n)=0.
\end{equation} 
And we have $|\phi^{-1}\rangle=\frac{\gamma\sqrt{2B(n+1)}}{E}(\hat{\Pi}|n+1\rangle)+\frac{W\beta}{\gamma^{2}+E\beta}(\hat{\Pi}|n\rangle)$. Besides, $\tilde{\mathcal{H}}_{n}^{(-1)}$ has a zero mode with $E=0$, the corresponding state is $(0, \hat{\Pi}_{E=0}|0\rangle)^{T}$. 

Comparing with the results of the effective-band picture, namely the effective Hamiltonian (\ref{effectiveH}) and quantization rules (\ref{quantizationrules}), we see that the cubic equation for $\tilde{\mathcal{H}}_{E}^{(-1)}$ is reproduced and the case for $n=0$ in (\ref{quantizationrules}) corresponds to its zero mode. Furthermore, after adding an extra state for $n=0$ in (\ref{effectiveH}) and (\ref{quantizationrules}), the equations for eigenvalues are exactly the same as the cubic equation (\ref{cubicplus}) for $\tilde{\mathcal{H}}_{E}^{(+1)}$. So the results of effective-band picture exactly reproduce the LL spectrum of the model. In the following we show that the LL spectrum obtained from the PDE method leads to the same results.

\subsection*{PDE solution for the case $\beta'=0$}

Here we solve the Hamiltonian without rotational invariance $\tilde{\mathcal{H}}_{k}^{(-1)}$ using partial differential equations (PDE) method. We now take $W$ to be a complex number and the Hamiltonian reads
\begin{equation}
\mathcal{H}=\left(\begin{array}{cc}
\beta(\tilde{k}_{x}^{2}+\tilde{k}_{y}^{2})-\beta k_{0}^{2}&\gamma(\tilde{k}_{x}+i\tilde{k}_{y})+W\\
\gamma(\tilde{k}_{x}-i\tilde{k}_{y})+W^{*}&0
\end{array}\right).
\end{equation}
In the Hamiltonian $\tilde{k}_{x,y}$ are understood as differential operators acting on wavefunctions $(\mathbf{y}(x,y),\mathbf{z}(x,y))^{T}$. The eigenvalue problem takes the following form
\begin{equation}
\mathcal{H}\left(\begin{array}{c}
\mathbf{y}(x,y)\\ \mathbf{z}(x,y)
\end{array}\right)=E\left(\begin{array}{c}
\mathbf{y}(x,y)\\ \mathbf{z}(x,y)
\end{array}\right),
\end{equation}
from which we get $\mathbf{z}=\frac{1}{E}[\gamma(\tilde{k}_{x}-i\tilde{k}_{y})+W^{*}]\mathbf{y}$. And we obtain an equation for $\mathbf{y}$,
\begin{eqnarray}
\begin{aligned}
\bigg[\big(\beta E+\gamma^{2}\big)(\tilde{k}_{x}^{2}+\tilde{k}_{y}^{2})+\gamma W(\tilde{k}_{x}-i\tilde{k}_{y})+\gamma W^{*}(\tilde{k}_{x}+i\tilde{k}_{y})+\big[|W|^{2}+\gamma^{2}B-E(E+\beta k_{0}^{2})\big]\bigg]\mathbf{y}=0.
\end{aligned}
\end{eqnarray}
To simplify the notation we define two constants, $\lambda_{1}=\beta E+\gamma^{2}$ and $\lambda_{2}=|W|^{2}+\gamma^{2}B-E(E+\beta k_{0}^{2})$.
The equation can be solved with the {\it Landau gauge}, in which we have
\begin{equation}
\bigg[\partial_{x}^{2}+\partial_{y}^{2}-B^{2}x^{2}-2iBx\partial_{y}+\frac{\gamma W}{\lambda_{1}}(\partial_{y}+i\partial_{x}-iBx)+\frac{\gamma W^{*}}{\lambda_{1}}(-\partial_{y}+i\partial_{x}+iBx)-\frac{\lambda_{2}}{\lambda_{1}}\bigg]\mathbf{y}=0.
\end{equation}
Because the PDE does not contain $y$ explicitly (only derivatives of $y$), one can separate the variables by defining $\mathbf{y}(x,y)=e^{ipy}u(x)$. We thus acquire an ODE for $u(x)$, which reads
\begin{equation}
\frac{d^{2}u(x)}{dx^{2}}+\frac{2i\gamma}{\lambda_{1}}(\text{Re}W)\frac{du(x)}{dx}-B^{2}\bigg[x-\frac{1}{B}(\frac{\gamma}{\lambda_{1}}(\text{Im}W)+p)\bigg]^{2}u(x)+\bigg(\frac{\gamma^{2}}{\lambda_{1}^{2}}(\text{Im}W)^{2}-\frac{\lambda_{2}}{\lambda_{1}}\bigg)u(x)=0.
\end{equation}
To further simplify the problem we define another function $u(x)=e^{-\frac{i\gamma\text{Re}W}{\lambda_{1}}x}\tilde{u}(x)$, and then introduce a shift of coordinate, $\tilde{x}=x-\frac{1}{B}(\frac{\gamma}{\lambda_{1}}(\text{Im}W)+p)$. After these, the ODE for $\tilde{u}$ is given by
\begin{equation}
\frac{d^{2}}{d\tilde{x}^{2}}\tilde{u}(\tilde{x})-B^{2}\tilde{x}^{2}\tilde{u}(\tilde{x})+\bigg[\frac{\gamma^{2}}{\lambda_{1}^{2}}|W|^{2}-\frac{\lambda_{2}}{\lambda_{1}}\bigg]\tilde{u}(\tilde{x})=0.
\end{equation}
Standard theory for 1D quantum harmonic oscillator gives that the ODE: $\frac{d^{2}\psi}{dx^{2}}-\frac{m^{2}\omega^{2}}{\hbar^{2}}x^{2}\psi+\frac{2mE}{\hbar^{2}}\psi=0$ has eigenvalues $E=(n+\frac{1}{2})\hbar\omega$. Comparing with current problem, we identify that the self-consistent equation for our ODE is 
\begin{equation}
\frac{\gamma^{2}}{\lambda_{1}^{2}}|W|^{2}-\frac{\lambda_{2}}{\lambda_{1}}=2B(n+\frac{1}{2}).
\end{equation}
The self-consistent equation for energy is therefore simplified to the following cubic equation with $n=0,1,2,\cdots$,
\begin{equation}
\beta E^{3}-\left(2\beta^{2}B(n+\frac{1}{2})-\gamma^{2}-\beta^{2}k_{0}^{2}\right)E^{2}-\left(\beta|W|^{2}+2\beta B\gamma^{2}(2n+\frac{3}{2})-\beta\gamma^{2}k_{0}^{2}\right)E-2B\gamma^{4}(n+1)=0.
\end{equation}
This is the same cubic equation as (\ref{Hminusonecubic}).

\subsection*{Numerical results for the case $\beta'\neq 0$}

In this section we study the model without rotational invariance with a non-vanishing $\beta'$, the Hamiltonian reads,
\begin{equation}
\mathcal{H}^{(\pm 1)}=\left(\begin{array}{cc}
\beta(k^{2}-k_{0}^{2})&\gamma k_{\mp}+W\\
\gamma k_{\pm}+W&-\beta'(k^{2}-k_{0}^{2})
\end{array}\right).
\end{equation}
Here we focus on $\mathcal{H}^{(+1)}$. Inside magnetic field $B$, the magnetic Hamiltonian can be written in terms of dimensionless quantities $\tilde{B}=\frac{2B}{k_{0}^{2}}$, $\tilde{\gamma}=\frac{\gamma}{\beta k_{0}}$, $\tilde{W}=\frac{W}{\beta k_{0}^{2}}$, and $\tilde{\beta}'=\frac{\beta'}{\beta}$, 
\begin{equation}
\label{Hplusone}
\mathcal{H}^{(+1)}=(\beta k_{0}^{2})\left(\begin{array}{cc}
\tilde{B}(a^{\dagger}a+\frac{1}{2})-1&\tilde{\gamma}\sqrt{\tilde{B}}a^{\dagger}+\tilde{W}\\
\tilde{\gamma}\sqrt{\tilde{B}}a+\tilde{W}&-\tilde{\beta}'[\tilde{B}(a^{\dagger}a+\frac{1}{2})-1]
\end{array}\right).
\end{equation}
We assume that the eigenstates take the form $\left(c+c_{0}|0\rangle+c_{1}|1\rangle+\cdots, d+d_{0}|0\rangle+d_{1}|1\rangle+\cdots\right)^{T}$, in which $c_{i}$ and $d_{i}$ are coefficients. Under the Hamiltonian (\ref{Hplusone}) these coefficients satisfy the following recurrence relations
\begin{eqnarray}
\begin{aligned}
c_{n}&\rightarrow [(n+\frac{1}{2})\tilde{B}-1]c_{n}+\tilde{\gamma}\sqrt{\tilde{B}}\sqrt{n}d_{n-1}+\tilde{W}d_{n},\\
d_{n}&\rightarrow -\tilde{\beta}'[(n+\frac{1}{2})\tilde{B}-1]d_{n}+\tilde{W}c_{n}+\tilde{\gamma}\sqrt{\tilde{B}}\sqrt{n+1}c_{n+1}
\end{aligned}
\end{eqnarray}
With the states arranged as $(c,d,c_{0},d_{0},c_{1},d_{1},\cdots, c_{n},d_{n},\cdots)^{T}$, the Hamiltonian matrix is given by
\begin{equation}
\label{Hmatrix}
\mathcal{H}^{(+1)}=\left(\begin{array}{cccccccc}
\frac{\tilde{B}}{2}-1&\tilde{W}&\cdots&&&&&\\
\tilde{W}&-\tilde{\beta}'(\frac{\tilde{B}}{2}-1)&&&&&&\\
&&\ddots&&&&&\\
&&&&&\vdots&&\\
&&&\tilde{\gamma}\sqrt{\tilde{B}n}&(n+\frac{1}{2})\tilde{B}-1&\tilde{W}&&\\
&&&&\tilde{W}&-\tilde{\beta}'[(n+\frac{1}{2})\tilde{B}-1]&\tilde{\gamma}\sqrt{\tilde{B}(n+1)}&\\
&&&&&\vdots&&\ddots
\end{array}\right).
\end{equation}
For numerical studies, we truncate the states to include only $c, c_{0}, c_{1},\cdots, c_{N-2}$ and $d, d_{0}, d_{1},\cdots, d_{N-2}$. Hence the Hamiltonian matrix (\ref{Hmatrix}) becomes a $2N\times 2N$ finite matrix and can be diagonalized. A test for $N=1000$ for the case $\beta'=0$ shows that the numerical LLs agree well with the exact solutions, especially for small LL indices. Also LLs near $E=0$ converge well for $N=1000$, so we take $N=1000$ for our numerical studies. 

\begin{figure}
\includegraphics[width=0.6\textwidth]{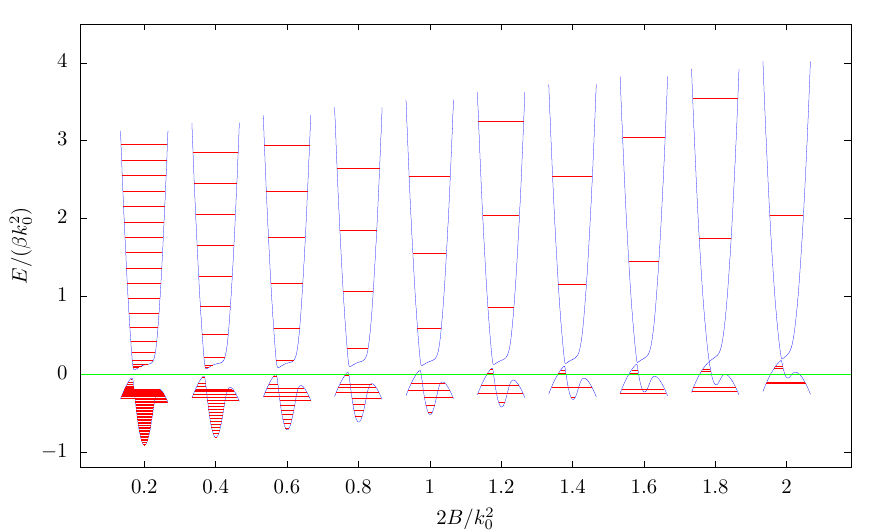}
\caption{Effective band for $\mathcal{H}^{(+1)}$ without rotation invariance ($k_{y}=0$), $\frac{\gamma}{\beta k_{0}}=0.2$, $\tilde{W}=\frac{W}{\beta k_{0}^{2}}=0.1$ and $\tilde{\beta}'=\frac{\beta'}{\beta}=0.1$. Imaginary chemical potential is set at $E=0$.}
\label{figEBB}
\end{figure}

Although quantization rules cannot be defined for $\beta'\neq 0$, we can still use generalized effective band from (\ref{effectiveH}) to enclose the LLs, in particular we take effective Hamiltonian 
\begin{equation}
\tilde{\mathcal{H}}_{E}^{(\pm 1)}=\left(\begin{array}{cc}
\beta (k^{2}-k_{0}^{2})\pm B\beta &\gamma k_{\mp}+W\\
\gamma k_{\pm}+W&-\beta' (k^{2}-k_{0}^{2})\pm B
\end{array}\right).
\end{equation}
In Fig. \ref{figEBB} we plot the effective band $\mathcal{H}^{(+1)}$ as a function of $k_{x}$ with $k_{y}=0$ with varying magnetic field, the LLs are obtained by numerically diagonalizing (\ref{Hmatrix}) ($N=1000$). The effective band can still show the evolution of LLs and capture the ALLs that appear in the original band.

\end{document}